\documentclass[aip,amsmath,amssymb,graphicx,reprint]{revtex4-1}
\usepackage{amsmath}
\usepackage{amssymb}
\usepackage{float}
\usepackage{tikz}
\usepackage{mathtools}
\usepackage{color,colortbl}
\usepackage{hyperref}
\usetikzlibrary{arrows}
\usepackage{caption}
\usepackage{blindtext}
\usepackage{subcaption}
\usepackage{graphicx}
\usepackage{dcolumn}
\usepackage{bm}
\usepackage[utf8]{inputenc}
\usepackage[T1]{fontenc}
\usepackage{mathptmx}
\usepackage{etoolbox}

\newcommand{\rhostar}{\ensuremath{{\rho_\star}}}
\newcommand{\psih}[1]{\ensuremath{{\hat{\psi}_{0 #1}}}}

\newcommand{\ee}[1]{\ensuremath{\mathcal{E}_{#1}}}
\newcommand{\vp}[1]{\ensuremath{v_{\parallel #1}}}
\newcommand{\vperp}[1]{\ensuremath{v_{\perp #1}}}
\newcommand{\ts}[1]{\ensuremath{{\theta^\star_{#1}}}}

\newcommand{\bstar}[1]{\ensuremath{B_{\parallel #1}^\star}}
\newcommand{\dint}[1]{\ensuremath{\,\text{d}#1}}

\newcommand{\ddt}[1]{\ensuremath{\frac{\text{d}{#1}}{\text{d}t}}}

\newcommand{\pp}[2]{\ensuremath{\frac{\partial #1}{\partial #2}}}
\newcommand{\fsa}[1]{\ensuremath{\left\langle #1 \right\rangle}}
\newcommand{\NL}{\ensuremath{\nonumber\\}}

\makeatletter
\def\@email#1#2{%
 \endgroup
 \patchcmd{\titleblock@produce}
  {\frontmatter@RRAPformat}
  {\frontmatter@RRAPformat{\produce@RRAP{*#1\href{mailto:#2}{#2}}}\frontmatter@RRAPformat}
  {}{}
}%
\makeatother
\begin{document}


\title[Gyrokinetic simulations with an evolving background]{Gyrokinetic flux-driven simulations in mixed TEM/ITG regime using a delta-f PIC scheme with evolving background}
\author{M. Murugappan}
 \email{moahan.murugappan@epfl.ch}

\author{L. Villard}

\author{S. Brunner}

\author{G. Di Giannatale}

\affiliation{Ecole Polytechnique F\'{e}d\'{e}rale de Lausanne (EPFL), Swiss Plasma Center (SPC), CH-1015 Lausanne, Switzerland%
}%

\author{B. F. McMillan}
\affiliation{CFSA, Department of Physics, University of Warwick, Coventry CV4 7AL, United Kingdom%
}%

\author{A. Bottino}
\affiliation{Max-Planck-Institut f\"{u}r Plasmaphysik, D-85748 Garching, Germany%
}%

\date{\today}

\begin{abstract}
In the context of global gyrokinetic simulations of turbulence using a Particle-In-Cell framework, verifying the delta-f assumption with a fixed background distribution becomes challenging when determining quasi-steady state profiles corresponding to given sources over long time scales, where plasma profiles can evolve significantly. The advantage of low relative sampling noise afforded by the delta-f scheme is shown to be retained by considering the background as a time-evolving Maxwellian with time-dependent density and temperature profiles. Implementation of this adaptive scheme to simulate electrostatic collisionless flux-driven turbulence in tokamak plasmas show small and non-increasing sampling noise levels, which would otherwise increase indefinitely with a stationary background scheme. The adaptive scheme furthermore allows one to reach numerically converged results of quasi-steady state with much lower marker numbers. 
\end{abstract}

\maketitle

\section{Introduction}

Magnetic fusion research relies heavily on its accurate modeling by computer simulations. In the most promising reactor configuration, the tokamak, the plasma is confined by magnetic fields in a toroidal vacuum chamber. A complete description of the plasma involves simulating regions of the core, edge, the Scrape-Off-Layer (SOL), and plasma-wall interaction.

To simulate fusion plasmas, many methods exist, which can be categorized by the physical assumptions made, dictated by the physical process of the plasma volume considered. This work focuses on the gyrokinetic Particle-In-Cell (PIC) method~\cite{Lin2007,Garbet2010,Ku2009,Parker1993}. The gyrokinetic formalism~\cite{Brizard2007,Hahm1988,Tronko2018} reduces the number of phase space variables from six to five, approximating the dynamics of plasma particle trajectories by gyrorings bound to evolving gyrocenters. The reduction in dynamics implies a time-scale separation between the fast cyclotron motion and the typical fluctuation time scales involved in turbulent processes. The PIC scheme relies on representing the evolution of particle distributions in phase space in terms of Lagrangian `markers', whose characteristics are dictated by the equations of motion. In this approach, the sources of the fields, i.e.~the charge and current densities, are obtained via Monte Carlo integration, thus giving the distribution a statistical interpretation~\cite{Aydemir1994}. This also means that it inherits the main problems of Monte Carlo sampling, notably the statistical sampling error problem referred to as `noise' in the following. Unless appropriate noise control measures are implemented, this problem severely limits the physically relevant simulation time. This is especially true for cases that exhibit significant deviation from initial profiles and/or large relative fluctuation amplitudes.

The aim of this work is to improve the statistical sampling noise problem of the PIC scheme in order to apply it to long transport time scale simulations, i.e. targeting cases with significant deviations from initial profiles. For plasma simulations where the distribution function $f$ of each species (ions, electrons possible impurities) does not deviate more than a few percent from its initial state $f_{\rm init}$, one usually adopts the delta-f splitting. The approach decomposes $f$ into a stationary (and often analytic) distribution $f_0$ and a time-dependent perturbation part $\delta f$, where only the latter is represented by numerical markers. This so-called delta-f PIC method is to be contrasted with the full-f PIC scheme, which represents the whole $f$ in terms of markers. The gain in noise reduction of the delta-f scheme relies on the reduced variance of the marker weights, provided that the assumption $\|\delta f\|/\|f\|\ll1$ for some definition of the norm $\|\cdot\|$ is met. However, for plasma simulations exhibiting a large perturbed component $\delta f$, i.e. such that $\|\delta f\|/\| f\|\sim 1$, one usually falls back to the full-f scheme, which entails using high marker numbers to achieve low enough noise levels. In order to still possibly retain the advantage of the delta-f scheme, one can also evolve~\cite{Allfrey2003,Brunner1999,Ku2016} $f_0$, albeit at a slower time scale than that of the fluctuating $\delta f$. This work explores the benefits of having such a time-evolving background by constraining $f_0$ to be a flux-surface-dependent Maxwellian which is time-dependent via its evolving gyrocenter density and temperature profiles. Another source of statistical sampling noise is related to `weight-spreading'~\cite{Chen1997,Brunner1999} resulting from the implementation of collision operators in the delta-f PIC scheme using a Langevin approach. However, this problem will not be addressed in this work as collisions are not considered.

Following the success of a previous work~\cite{Murugappan2022} by the author using a similar approach in a simplified setup (sheared slab geometry, adiabatic electrons), the adaptive scheme is implemented in tokamak relevant axisymmetric toroidal geometry in the frame of the global gyrokinetic PIC code ORB5~\cite{Lanti2020}. The simulations are electrostatic and `flux-driven', with gyrokinetic ions while electrons have a hybrid response~\cite{Lanti2018} and instabilities being driven by both ions and electrons in a mixed Ion-Temperature-Gradient-Trapped-Electron-Mode (ITG-TEM) regime. Background density and temperature adaptations are made for both species independently. When compared to non-adaptive cases, results from the adaptive scheme exhibit low errors resulting from statistical sampling noise. Evolving the plasma profiles in the presence of radially localized sources up to their quasi-steady state using the adaptive scheme will be shown to require much lower marker numbers than with the non-adaptive scheme.

This paper is organized as follows. Sec. \ref{sec:physical} introduces the physical model to be solved. Namely, the Vlasov-Maxwell equation in the first order gyrokinetic approximation. A discussion of the Quasi-Neutrality-Equation (QNE) then follows, elaborating on the electrons' hybrid response. It concludes by describing the functional form of the different possible source terms used in this work. Sec. \ref{sec:numerical} elaborates on the use of a control variate in the form of either a canonical or local Maxwellian, with flux-surface-averaged (f.s.a.) density, parallel flow and temperature having an explicit time-dependence. The relaxation equations that connect the three lowest order velocity moments of $f_0$ and $\delta f$ are then explained. The modification of $\delta f$ requires a correction term to be added to the right-hand-side (r.h.s.) of the QNE. Sec. \ref{sec:profiles} details the considered initial profiles of density and temperature, along with the form of the actual heat sources for each species used in this work. Information about grid resolution, Fourier filtering, source strengths, and adaptive scheme parameters are given. Sec. \ref{sec:time} begins the result section with a discussion on time traces of various transport parameters and sampling noise diagnostics, comparing them between non-adaptive and adaptive cases. Sec. \ref{sec:early} investigates the effect of the adaptive scheme on time-averaged profiles of f.s.a.~density and temperature. This section also examines the need of the QNE r.h.s.~correction. Sec. \ref{sec:pvol} diagnoses the time evolution of sampled phase-space volume and marker distribution, which reveals in some simulation cases a problem of under-sampling. Sec. \ref{sec:late} discusses the f.s.a.~profiles of density and temperature, time-averaged over the  quasi-steady state, results of which could be acquired only under the adaptive scheme. Sec. \ref{sec:conclusion} then summarizes the main merits of the adaptive scheme demonstrated in this work, along with possible future developments and applications directions. 

\section{Physical model}\label{sec:physical}

All simulations carried out in the framework of this work only consider a singly-charged ($Z=1$) ion species and electrons. The distribution $f_{\rm j}$ of the j$^{\rm th}$ species is governed by the gyrokinetic equation
\begin{eqnarray}
    \ddt{f_{\rm j}} &=& \left(\pp{}{t} + \ddt{\vec{R}}\cdot\nabla_{\vec{R}}+\ddt{\vp{}}\pp{}{\vp{}}+\ddt{\mu}\pp{}{\mu}\right)f_{\rm j} = S_{\rm j}, \label{eq:vlasov}
\end{eqnarray}
where $S_{\rm j}$ is a general source term. Here, $f_{\rm j}$ is the gyrocenter distribution in $5$D gyrocenter phase space $\vec{Z}=[\vec{R},\vp{},\mu]$, where $\vec{R}$ is the $3$D configuration space vector, $\vp{}$ the parallel velocity and $\mu=\vperp{}^2/2B$ the magnetic moment per mass, with $\vperp{}$ the perpendicular velocity and $B$ the strength of the local magnetic field $\vec{B}=B\hat{b}$. ORB5 uses magnetic coordinates $[s,\ts{},\varphi]$ for representing the gyrocenter position $\vec{R}$ in tokamak geometry, with $s=\sqrt{\psi/\psi_{\rm edge}}$ the normalized radial coordinate expressed in terms of the poloidal magnetic flux function $\psi$ and its value at the edge of the radial boundary $\psi_{\rm edge}$. $\ts{}$ is the straight-field-line poloidal angle, and $\varphi$ is the toroidal angle. As this work is concerned with electrostatic turbulence, the equations of motion of $\vec{Z}$ are given by
\begin{eqnarray}
    \ddt{\vec{R}} &=& \vp{}\hat{b} + \frac{1}{\bstar{}}\left[\frac{\mu B+\vp{}^2}{\Omega_{\rm c}}\hat{b}\times\nabla B - \frac{\vp{}^2}{\Omega_{\rm c}}\hat{b}\times(\hat{b}\times\nabla\times\vec{B})- \right. \NL
    & & \left.\nabla\tilde{\phi}\times\hat{b}\right] \NL
    \ddt{\vp{}} &=& -\mu\nabla B\cdot\left[\hat{b}-\frac{\vp{}}{\bstar{}\Omega_{\rm c}}\hat{b}\times(\hat{b}\times\nabla\times\vec{B})\right] - \NL
    & & \frac{\nabla\tilde{\phi}}{\Omega_{\rm c}}\cdot\left\{\vec{B} + \frac{B\vp{}}{\bstar{}\Omega_{\rm c}}[\hat{b}\times\nabla B-\hat{b}\times(\hat{b}\times\nabla\times\vec{B})]\right\}\NL
    \ddt{\mu} &=& 0. \label{eq:eom}
\end{eqnarray}
Here, $\Omega_{\rm c}=qB/m$ is the species cyclotron frequency with mass $m$ and electric charge $q$, $\bstar{}=B(1+\vp{}\hat{b}\cdot\nabla\times\hat{b}/\Omega_{\rm c}$), and $\tilde{\phi}$ is the gyroaveraged electrostatic field $\phi$ evaluated at the gyrocenter position, given by
\begin{eqnarray}
    \tilde{\phi}(\vec{R},\mu,t) &=& \frac{1}{2\pi}\int_0^{2\pi}\dint{\alpha}\, \phi[\vec{R}+\vec{\rho}_L(\mu,\alpha),t], \nonumber
\end{eqnarray}
with $\alpha$ the gyrophase and $\vec{\rho}_{\rm L}$ the local Larmor radius vector with amplitude $\rho_{\rm L}=\sqrt{2m\mu B}/|\Omega_{\rm c}|$. The set of Eqs.(\ref{eq:eom}) is nonlinear as it depends on the self-consistent field $\phi(\vec{r},t)$ satisfying the QNE,
\begin{eqnarray}
     -\nabla_\perp\cdot\left(\frac{m_in_{0i}(\psi)}{eB^2}\nabla_\perp\phi\right) &=& \int\frac{\dint{\alpha}}{2\pi}\int\dint{\Omega}\, {f_i\,\delta[\vec{R}+\vec{\rho}_{\rm L}-\vec{r}]} - \NL
    & & \int\dint{\Omega}\, {f_e\,\delta[\vec{R}-\vec{r}]}. \label{eq:qne}
\end{eqnarray}
Here, the subscripts $i$ and $e$ denote ion and electron quantities respectively, and $\dint{\Omega}=J_Z(\vec{Z})\dint{^3R}\dint{\vp{}}\dint{\mu}$ is the phase space differential volume element with Jacobian $J_Z$, with $\dint{^3R}$ the configuration space volume differential element. The left-hand-side (l.h.s.) of Eq.(\ref{eq:qne}) represents the linearized ion polarisation density in the long wavelength limit $k_\perp\rho_{{\rm th}i}\ll1$, with $k_\perp$ the perpendicular wavelength of the turbulence and $\rho_{{\rm th}i}=v_{{\rm th}i}/\Omega_{{\rm c}i}$ the ion thermal Larmor radius with thermal velocity $v_{{\rm th}i}=\sqrt{T_i/m_i}$. The linearization is due to the $\delta f$ splitting of the distribution function
\begin{eqnarray}
    f_{\rm j} &=& f_{\rm 0j} + \delta f_{\rm j} \label{eq:deltaf}
\end{eqnarray}
into the background $f_0$ and $\delta f$ components. In Eq.(\ref{eq:qne}), $n_{0i}$ represents the ion background guiding center density, given by $n_{0i}=\int\dint{^3v}\,f_{0i}$ with $\dint{^3v}=2\pi\bstar{}\dint{\vp{}}\dint{\mu}$, and is taken to be a flux function. Finally, the perpendicular gradient operator $\nabla_\perp$ is approximated to lie in the poloidal plane, i.e.~$\nabla_\perp\approx\nabla s\pp{}{s}+\nabla\ts{}\pp{}{\ts{}}$. The r.h.s.~of Eq.(\ref{eq:qne}) represents the difference between the ion gyrodensity and the electron density, where the drift-kinetic assumption for the electron applies. We further assume that the background ion gyrodensity perfectly cancels the background electron density, effectively replacing $f_{\rm j}$ with $\delta f_{\rm j}$ on the r.h.s. of Eq.(\ref{eq:qne}).

In order to simulate Trapped-Electron-Modes (TEMs), but yet avoid having to possibly resolve Electron-Temperature-Gradient (ETG) modes, this work uses the upgraded hybrid electron response model~\cite{Idomura2016,Lanti2018}. With this approximate model, the QNE reads
\begin{eqnarray}
    & & \alpha_{\rm P}(\psi,\ts{})\frac{en_{0e}(\psi)}{T_{0e}(\psi)}(\phi-\fsa{\phi}) -\nabla_\perp\cdot\left(\frac{m_in_{0i}(\psi)}{eB^2}\nabla_\perp\phi\right) \NL
    &=& \int\dint{\Omega}\, \int\frac{\dint{\alpha}}{2\pi}\, \delta[\vec{R}+\vec{\rho}_{\rm L}-\vec{r}]\, \delta f_i - \NL
    & & \int\dint{\Omega}\, \delta[\vec{R}-\vec{r}]\, \delta f_{{\rm T}e} - \left.\int\dint{\Omega}\, \delta[\vec{R}-\vec{r}]\, \delta f_{{\rm P}e}\right|_{(m,n)=(0,0)}. \NL
    & & \label{eq:qne_hyb}
\end{eqnarray}
The first term on the l.h.s.~of Eq.(\ref{eq:qne_hyb}) represents the adiabatic response of the passing electrons, with passing fraction $\alpha_{\rm P}(\psi,\ts{}) = 1-\sqrt{1-B(\psi,\ts{})/B_{\rm max}(\psi)}$, where $B_{\rm max}(\psi)$ is the maximum amplitude of $B$ on the flux surface $\psi$. $T_{0e}$ is the electron background temperature, also a flux function, and $\fsa{\phi}$ is the f.s.a.~potential, given by
\begin{eqnarray}
    \fsa{\phi}(s,t) &=& \frac{\int\dint{\ts{}}\dint{\varphi}\,J_s\,\phi(s,\ts{},\varphi,t)}{\int\dint{\ts{}}\dint{\varphi}\,J_s}, \label{eq:fsa}
\end{eqnarray}
with $J_s(s,\ts{})$ the configuration space Jacobian. The second and third terms on the r.h.s.~of Eq.(\ref{eq:qne_hyb}) represent the perturbed densities of the trapped (subscript T) and passing (subscript P) electron, respectively, which are treated drift-kinetically. Note that it is only the zonal component $(m,n)=(0,0)$ of the passing electrons that gives a drift-kinetic response, where $m$ and $n$ are the poloidal and toroidal mode numbers. This electron model ensures ambipolarity and correctly captures the Geodesic Acoustic Mode (GAM) frequency and damping rate~\cite{Lanti2018}. 

In order to reach quasi-steady state, it was shown~\cite{Krommes1999} that some form of dissipation is required. Given that we neglect physical collisions, we therefore introduce a noise control Krook operator $S_{\rm nj} =-\gamma_{\rm n}(f_{\rm j}-f_{\rm 0j}) + S_{\rm nj}^{\rm c}$ of rate $\gamma_{\rm n}$, which is species independent and is taken to be a small fraction (typically around $10\%$) of the maximum linear growth rate of the instabilities driving the turbulence. $S_{\rm nj}$ relaxes the distribution $f_{\rm j}$ to the background $f_{\rm 0j}$, while conserving f.s.a.~density, parallel moment, residual zonal flows, and energy. These conservation properties are ensured by the correction term $S_{\rm nj}^{\rm c}$. Besides serving as a noise control operator for the PIC scheme, temperature-gradient-driven simulations of this work use the source term $S_{\rm nj}$ as a fixed power heat source by imposing conservation of all f.s.a.~moments except kinetic energy. For flux-driven simulations, i.e. with fixed prescribed power heat source, the f.s.a.~heating operator used is given by
\begin{eqnarray}
    S_{\rm Hj}(\psi,\ee{}) &=& \gamma_{\rm Hj}G_{\rm Hj}(\psi)\pp{f_{\rm Lj}}{T_{\rm Hj}} + S_{\rm Hj}^{\rm c} \label{eq:heat}
\end{eqnarray}
with local Maxwellian
\begin{eqnarray}
    f_{\rm L}(\psi,\ee{}) &=& \frac{n_{\rm H}(\psi)}{[2\pi T_{\rm H}(\psi)/m]^{3/2}}\exp\left(-\frac{\ee{}}{T_{\rm H}(\psi)/m}\right). \label{eq:heat_maxwellian}
\end{eqnarray}
Here, $\ee{}=v^2/2=\vp{}^2/2+\mu B$ is the kinetic energy per mass of the gyrocenter. The heat source $S_{\rm H}$ is parameterized by the f.s.a.~profiles $n_{\rm H}$ and $T_{\rm H}$ appearing in Eq.(\ref{eq:heat_maxwellian}), which in this work are taken to be the initial $n_0$ and $T_0$ profiles of each species. $\gamma_{\rm H}$ and $G_{\rm H}$ of Eq.(\ref{eq:heat}) represent the heating rate and radial heating profile normalized to the unit of temperature, respectively. The form $\pp{f_{\rm L}}{T_{\rm H}}$ analytically ensures no f.s.a.~density or momentum source. Nonetheless, $S_{\rm H}^{\rm c}$ ensures numerical conservation of f.s.a.~density, parallel momentum, and residual zonal flows to round-off. Finally, to damp turbulence at the outer radial edge ($s=1$) of the simulation domain, a buffer in the form of a Krook operator $S_{\rm bj}(s)=-\gamma_{\rm b}[(s-s_{\rm b})/(1-s_{b})]^4\delta f_j$, of species-independent damping rate $\gamma_{\rm b}$ is used. $S_{\rm b}$ is parametrized by the species-dependent buffer entrance $s=s_{\rm b}$ and only acts in the radial range $s\in[s_{\rm b},1]$. Taken together, the r.h.s. of Eq.(\ref{eq:vlasov}) now reads
\begin{eqnarray}
    S_{\rm j} &=& S_{\rm nj} + S_{\rm nj}^{\rm c} + S_{\rm Hj} + S_{\rm Hj}^{\rm c} + S_{\rm bj}. \nonumber
\end{eqnarray}
The explicit expressions of correction terms $S_{\rm n}^{\rm c}$ and $S_{\rm H}^{\rm c}$ can be found in Refs.~\cite{Villard2019,Murugappan2022}. We shall henceforth drop the $j^{\rm th}$ species index for ease of notation.

\section{Numerical methods} \label{sec:numerical}

Under the delta-f splitting defined by Eq.(\ref{eq:deltaf}), we choose $f_0=f_{\rm eq}$, where $f_{\rm eq}$ is an equilibrium distribution of the unperturbed system in the absence of the heat source term $S_{\rm H}$. This implies that $f_{\rm eq}$ has to be a function of the constants of motion of the unperturbed system. Namely, the modified canonical toroidal momentum~\cite{Angelino2006} $\psih{}=\psi + F(\psi)\vp{}/\Omega_{\rm c} + \psi_{0corr}$, the kinetic energy $\ee{}$, and the magnetic moment $\mu$. Here, $F(\psi)$ is the poloidal current flux function, as it appears in the representation of the axisymmetric equilibrium magnetic field $\vec{B}=F(\psi)\nabla\varphi+\nabla\psi\times\nabla\varphi$, and $\psi_{0corr}$ is a correction term that results in $\psih{}\sim\psi$ on average along unperturbed trajectories. Under the Monte-Carlo interpretation~\cite{Aydemir1994}, $f_0$ serves as a control variate, provided that the delta-f assumption
\begin{eqnarray}
    \frac{\|\delta f\|}{\|f\|}\ll1, \label{eq:assumption}
\end{eqnarray}
holds. However, when profiles evolve secularly over long simulation times, this assumption may not hold and in that case $f_0$ fails to be an optimal control variate that reduces noise. This motivates the introduction of an explicit time dependence in $f_0$. Thus, the control variate used in this work takes the form of a `canonical Maxwellian'~\cite{Angelino2006}:
\begin{eqnarray}
    f_0(\vec{Z},t) &=& \frac{n_0(\psih{},t)}{[2\pi T_0(\psih{},t)]^{3/2}}\exp\left[-\frac{[\vp{}-u_0(\psih{},t)]^2/2+\mu B}{T_0(\psih{},t)/m}\right]. \NL
    & & \label{eq:f0cm}
\end{eqnarray}
Here, $u_0$ is the background parallel flow whose radial coordinate indexed by $\psih{}$, with assumption $u_0(\psih{},t=0)=0$. Therefore, at initial time, $f_0(\vec{Z},t=0) = f_0(\psih{},\ee{},\mu,t=0) = f_{\rm eq}$. An alternative choice of control variate, which is not $f_{\rm eq}$, is given by the `local Maxwellian'
\begin{eqnarray}
    f_0(\vec{Z},t) &=& \frac{n_0(\psi,t)}{[2\pi T_0(\psi,t)]^{3/2}}\exp\left[-\frac{[\vp{}-u_0(\psi,t)]^2/2+\mu B}{T_0(\psi,t)/m}\right]. \NL
    & & \label{eq:f0lm}
\end{eqnarray}
Note that time dependence enters the control variate of Eq.(\ref{eq:f0lm}) via the f.s.a.~background profiles of density $n_0$, parallel flow $u_0$ and temperature $T_0$ of the gyrocenters. In the following, Eq.(\ref{eq:f0lm}) will be used to develop the time dependent control variate scheme.

As the splitting given by Eq.(\ref{eq:deltaf}) does not uniquely set $f_0$, we make use of this flexibility and consider, for each species, time evolution equations for $n_0$, $u_0$ and $T_0$ which progressively feed the contributions of the corresponding three velocity moments from the fluctuating part $\delta f$ into the control variate $f_0$:
\begin{eqnarray}
    \pp{n_0}{t}(\psi,t) &=& \alpha_n\fsa{\int\dint{^3v}\, \delta f}, \label{eq:relax_dens} \\
    \pp{}{t}\{n_0u_0\}(\psi,t) &=& \alpha_u\fsa{\int\dint{^3v}\, \vp{}\, \delta f}, \label{eq:relax_nvpar} \\
    \pp{}{t}\left\{\frac{3n_0T_0}{2m}+\frac{n_0u_0^2}{2}\right\}(\psi,t) &=& \alpha_E\fsa{\int\dint{^3v}\, \ee{}\, \delta f}. \label{eq:relax_ekin}
\end{eqnarray}
Here, $\alpha_n$, $\alpha_u$ and $\alpha_E$ are user-defined constants, representing adaptation rates for the background density $n_0$, parallel momentum $n_0u_0$ and kinetic energy per mass $E_{\rm kin0}=3n_0T_0/2m + n_0u_0^2/2$ for the background $f_0$. Note that Eqs.(\ref{eq:relax_dens})-(\ref{eq:relax_ekin}) are ad-hoc, and in particular, do not correspond exactly to taking the respective velocity moments of $f_0$ of Eq.(\ref{eq:f0lm}) due the negligence of $\vp{}$-dependence in the velocity Jacobian $2\pi\bstar{}$.

The background profiles of $n_0$, $u_0$ and $T_0$ are periodically updated by solving Eqs.(\ref{eq:relax_dens})-(\ref{eq:relax_ekin}) via an explicit forward Euler scheme, with the r.h.s.~of Eqs.(\ref{eq:relax_dens})-(\ref{eq:relax_ekin}) actually being time-averaged within the elapsed period of length $N_t\Delta t$, where $\Delta t$ is the marker integration time step and $N_t$ is a user-defined integer. It is found that $\alpha N_t\Delta t<1$, for $\alpha\in[\alpha_n,\alpha_u,\alpha_E]$, give numerically stable results. Finally, if one chooses to use Eq.(\ref{eq:f0cm}) as the time dependent control variate, after time-stepping Eqs.(\ref{eq:relax_dens})-(\ref{eq:relax_ekin}) the following assignment is made:
\begin{eqnarray}
    n_0(\psih{},t) &\leftarrow& n_0(\psi,t), \NL
    u_0(\psih{},t) &\leftarrow& u_0(\psi,t), \NL
    T_0(\psih{},t) &\leftarrow& T_0(\psi,t). \nonumber
\end{eqnarray}
This introduces slight modifications to the f.s.a.~background profiles $n_0(\psi,t)$, $u_0(\psi,t)$ and $T_0(\psi,t)$ due to the $\vp{}$-dependence in $\psih{}$. For simplicity, in this work we did not carry out adaptation of the background parallel flow, and therefore set $\alpha_u=0$. Thus, $u_0=0$ at all times and $f_0$ of Eq.(\ref{eq:f0cm}) is always $f_0(t)=f_{\rm eq}$, i.e. a gyrokinetic equilibrium of the unperturbed system with $n_0(t)$ and $T_0(t)$.

Having applied Eqs.(\ref{eq:relax_dens})-(\ref{eq:relax_ekin}), Eq.(\ref{eq:deltaf})
implies $f_0'+\delta f' = f_0+\delta f$, where primed and unprimed notations refer respectively quantities immediately prior and after the background adaptation. In order to keep the r.h.s. of Eq.(\ref{eq:qne_hyb}) invariant to adaptation of background profiles, the correction term
\begin{eqnarray}
    I &=& \int\dint{\Omega}\int\frac{\dint{\alpha}}{2\pi}\,\delta[\vec{R}+\vec{\rho}_{\rm L}-\vec{r}]\, (f_{0i}'-f_{0i}) - \NL
    & & \int\dint{\Omega}\,\delta[\vec{R}-\vec{r}]\, (f_{0e}'-f_{0e}) \NL
    & & \label{eq:qne_corr}
\end{eqnarray}
must be appended. Given the analytic form for $f_0$ from Eq.(\ref{eq:f0lm}), Eq.(\ref{eq:qne_corr}) is calculated using quadratures on a 5D phase space grid. An alternative, and computationally cheaper way to avoid appending the correction term is to equate the two terms of Eq.(\ref{eq:qne_corr}). This means that the change in the background electron density is set to be approximately equal to the change in the background ion gyrodensity, i.e.
\begin{eqnarray}
    n_{0e}' - n_{0e} &=& \fsa{\int\dint{\Omega}\int\frac{\dint{\alpha}}{2\pi}\,\delta[\vec{R}+\vec{\rho}_{\rm L}-\vec{r}]\, (f_{0i}'-f_{0i})}, \NL
    & & \label{eq:qnrhs_fsa}
\end{eqnarray}
where the f.s.a. is defined as in Eq.(\ref{eq:fsa}). Equation (\ref{eq:qnrhs_fsa}) satisfies the quasineutrality of the backgrounds only approximately, because of the difference between the ion density and the f.s.a.~ion gyrodensity.

\section{Profiles and simulation parameters} \label{sec:profiles}

\begin{figure}[H]
	\centering
	\includegraphics[width=0.8\linewidth]{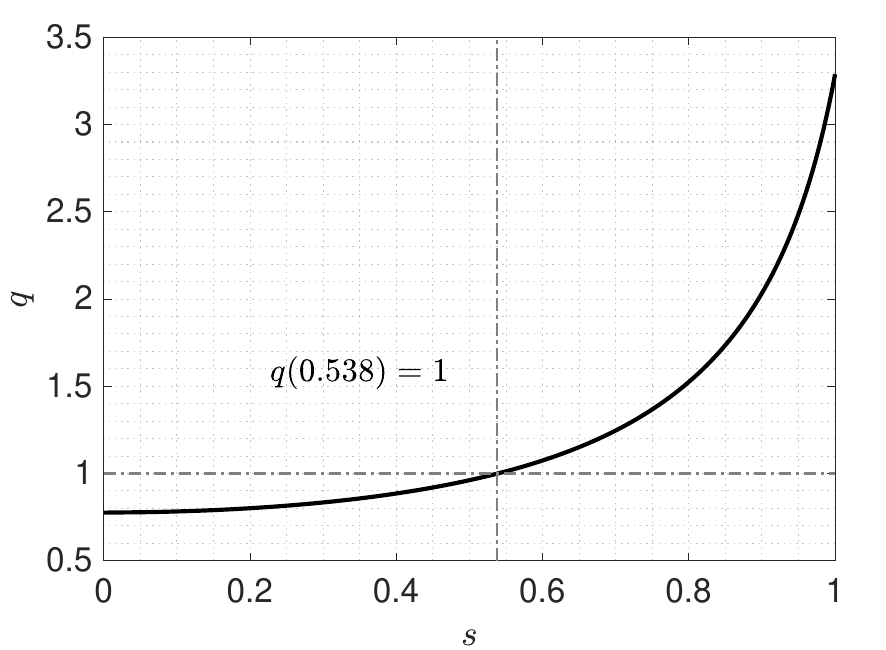}
	\caption{Safety factor profile for the TCV shot \#43516 used in this work.}
	\label{fig:q}
\end{figure}

For all simulations of this work, the ideal MHD equilibrium is computed by the CHEASE code~\cite{Luetjens2006} based on the TCV shot \#43516. It has an aspect ratio of $3.64$, an elongation of $1.44$, and a triangularity of $0.20$ at the last closed flux surface. The safety factor $q(s)$ is shown in Fig.\ref{fig:q}. The $q=1$ flux surface is located at $s=0.538$. The reference magnetic surface taken for normalization is at $s_0=1$. At the outer edge $\rhostar(s_0)=\rho_s(s_0)/a=1/245$, where $\rho_s(s)=\sqrt{T_{0e}(s)/m_i}$ is the ion sound Larmor radius, and $a$ the minor radius. The profile for the background ion and electron density and temperature is described by the following functional form, designated by $G$:
\begin{eqnarray}
    G(\rho_V) &=& \begin{cases}
        g_0 + g_2\rho_V^2 &\text{for} \hspace{1mm} 0\le\rho_V\le\rho_{\rm core} \\
        G_{\rm ped}\exp[-\kappa_T(\rho_V-\rho_{\rm ped})] &\text{for} \hspace{1mm}\rho_{\rm core}<\rho_V\le\rho_{\rm ped} \\
        G_1+\mu_G(1-\rho_V) &\text{for} \hspace{1mm} \rho_{\rm ped}<\rho_V\le1
    \end{cases}, \NL
    & & \label{eq:profile}
\end{eqnarray}
where $\rho_V$ is the radial coordinate $\rho_V(\psi)=\sqrt{V(\psi)/V(\psi_{\rm edge})}$, with $V(\psi)$ the volume enclosed by the flux surface label $\psi$. $g_0$ and $g_2$ are coefficients determined such that $G$ and $\dint{G}/\dint{\rho_V}$ are continuous at $\rho_V=\rho_{\rm core}$, and $G_{\rm ped}=G_1+\mu_G(1-\rho_{\rm ped})$. The functional form, was found to describe well the experimentally observed profiles in TCV L-mode discharges~\cite{Sauter2014}. The values of the parameters in Eq.(\ref{eq:profile}) for the ion and electron density and temperature profiles are shown in Tab.~\ref{tab:profile_params}. All densities and temperatures are normalized by $\bar{n}=N_{\rm ph}/V(\psi_{\rm edge})$ and $T_{0e}(s_0)$ respectively. Here, $N_{\rm ph}$ is the physical total number of particles.

\begin{table}[h]
    \centering
    \begin{tabular}{|c|c|c|c|c|}
    \hline
      & \multicolumn{2}{|c|}{Ions} & \multicolumn{2}{c|}{Electrons} \\
    \hline
    Parameter & Density & Temperature & Density & Temperature\\
    \hline
    $\rho_{\rm core}$ & $0.4016$ & $0.4016$ & $0.4016$ & $0.4016$ \\
    $\rho_{\rm ped}$ & $0.8$ & $0.8$ & $0.8$ & $0.8$ \\
    $\kappa$ & $2.3$ & $2.3$ & $2.3$ & $2.5$ \\
    $\mu$ & $5.0$ & $6.0$ & $5.0$ & $10.0$ \\
    $G_1$ & $1.0$ & $1.0$ & $1.0$ & $1.0$ \\
    \hline
    \end{tabular}
    \caption{Profile parameters as defined in Eq.(\ref{eq:profile}) for initial density and temperature for both background ions and electrons.}
    \label{tab:profile_params}
\end{table}

\begin{figure}[H]
	\centering
	\begin{subfigure}[c]{0.49\columnwidth}
	    \caption{Ions \label{fig:sourcei}}
	    \centering
		\includegraphics[width=1.0\linewidth]{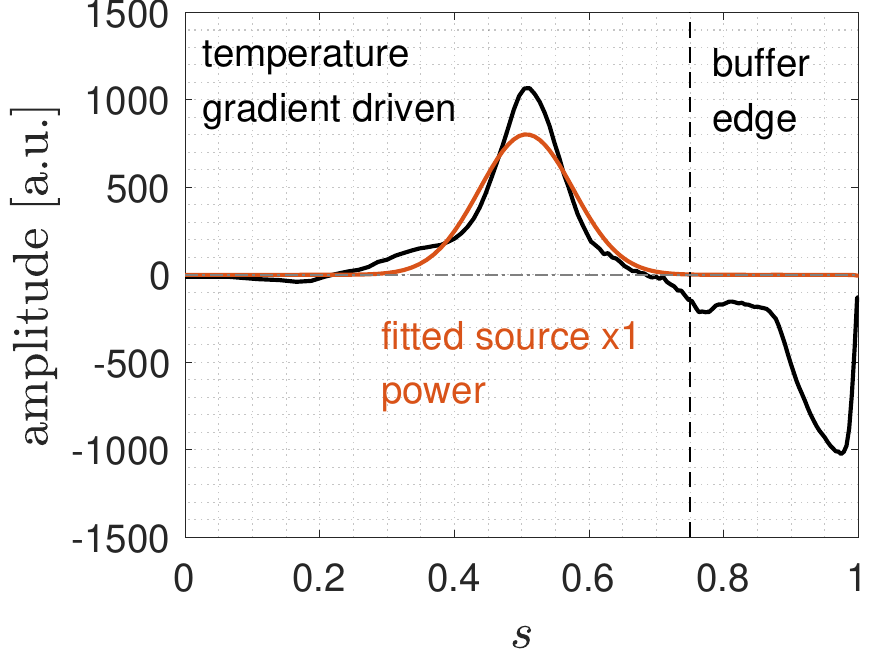}
	\end{subfigure}
	\begin{subfigure}[c]{0.49\columnwidth}
	    \caption{Electrons \label{fig:sourcee}}
	    \centering
		\includegraphics[width=1.0\linewidth]{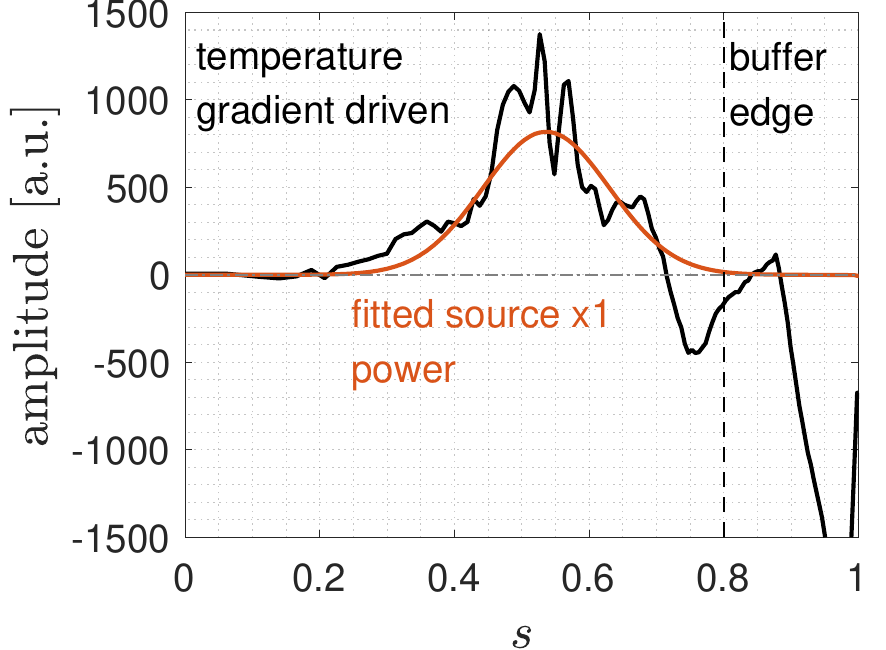}
	\end{subfigure}
	\caption{Fixed heat source profiles $G_H(\psi)$ (orange) for the flux-driven simulations fitted to the effective flux surface and time-averaged heat source of the temperature-gradient-driven run using the non-adaptive case with $N_p=256$M. Buffer edges for the ions and electrons are taken to be $s_b=0.75$ and $s_b=0.80$, respectively. Both species have a common buffer strength $\gamma_{\rm b}=11\%\gamma_{\rm max}$.}
	\label{fig:source}
\end{figure}

The grid resolution for the radial $s$, poloidal $\ts{}$ and toroidal $\varphi$ is taken to be $N_s\times N_{\ts{}}\times N_\varphi=256\times512\times256$, where $N$ represents the number of intervals. Toroidal Fourier modes in the range of $0\le n\le 64$ will be simulated, only keeping poloidal Fourier modes $m\in[nq(s)-\Delta m,nq(s)+\Delta m]$ with half-width $\Delta m=5$. To lighten the computational cost `heavy' electrons are considered, i.e. with an electron-ion mass ratio $m_e/m_i=1/200$. The time resolution is $c_s\Delta t/a=4.1\times10^{-3}$, and the maximum linear growth rate is found to be $\gamma_{\rm max}a/c_s=0.931$. The strength of the Krook operator $S_{\rm n}$ for noise control is set to $\gamma_{\rm n}=11\%\gamma_{\rm max}$. For the flux-driven simulations considered here, the radial heat source profile $G_{\rm H}(\psi)$ of Eq.(\ref{eq:heat}) is approximated by fitting a Gaussian function around the peak of the time-averaged effective heat source at quasi-steady state of a previously run temperature-gradient-driven simulation with the initial profiles given by Tab.~\ref{tab:profile_params}. The heat source strength $\gamma_{\rm H}$ is determined by equating the integrated power to that of the effective heat source. The heat sink at the radial edge is replaced by the Krook buffer $S_b$. The calculation for the heat source is done for both ions and electrons separately. An example of the heat source profiles for ions and electrons are shown in Fig.\ref{fig:source}.

Simulations are initialized by loading markers in phase space and initializing their weights using quasi-random distributions. All adaptive cases have adaptation rates $\alpha_n=\alpha_E=11\%\gamma_{\rm max}$ and adaptation period set with $N_t=100$ for both ions and electrons. Even when considering the hybrid electron model, adapted electron profiles include contributions from both passing and trapped electrons. Unless stated otherwise, all time traces have a moving time averaging window of $c_s\Delta t/a=4.085$. Marker numbers $N_p$ are displayed in millions (M). Comparison will be made between the non-adaptive (standard) case with $N_p=256$M, $128$M and the adaptive case with $N_p=128$M, $64$M. The choice of simulations with different $N_p$ allows for the discerning of sampling noise's effect on different diagnostic measures.

\section{Results}
\subsection{Time traces} \label{sec:time}

We define an effective heat diffusivity $\chi$ from the f.s.a.~radial component of the heat flux $\vec{q}_H$ resulting from gyrocenter $E\times B$ drifts. $\vec{q}_H$ is composed of the contributions from the kinetic energy $\vec{q}_{\rm kin}$, the field potential $\vec{q}_{\rm pot}$, and the particle flux $\vec{\Gamma}$, all of which are given respectively by~\cite{Vernay2010}
\begin{eqnarray}
    \vec{q}_{\rm kin} &=& \int\dint{^3v}\, m\ee{} \frac{-\nabla\tilde{\phi}\times\vec{B}}{B\bstar{}}\delta f, \NL
    \vec{q}_{\rm pot} &=& q\int\dint{^3v}\, \phi \frac{-\nabla\tilde{\phi}\times\vec{B}}{B\bstar{}}\delta f, \NL
    \vec{\Gamma} &=& \int\dint{^3v}\, \frac{-\nabla\tilde{\phi}\times\vec{B}}{B\bstar{}}\delta f, \NL
    \vec{q}_{\rm H} &=& \vec{q}_{\rm kin} + \vec{q}_{\rm pot} - \frac{5}{2}mv_{\rm th}^2\vec{\Gamma}. \nonumber
\end{eqnarray}

\begin{figure}[H]
	\centering
	\begin{subfigure}[c]{1.0\columnwidth}
	    \caption{\label{fig:chii}}
	    \centering
		\includegraphics[width=1.0\linewidth]{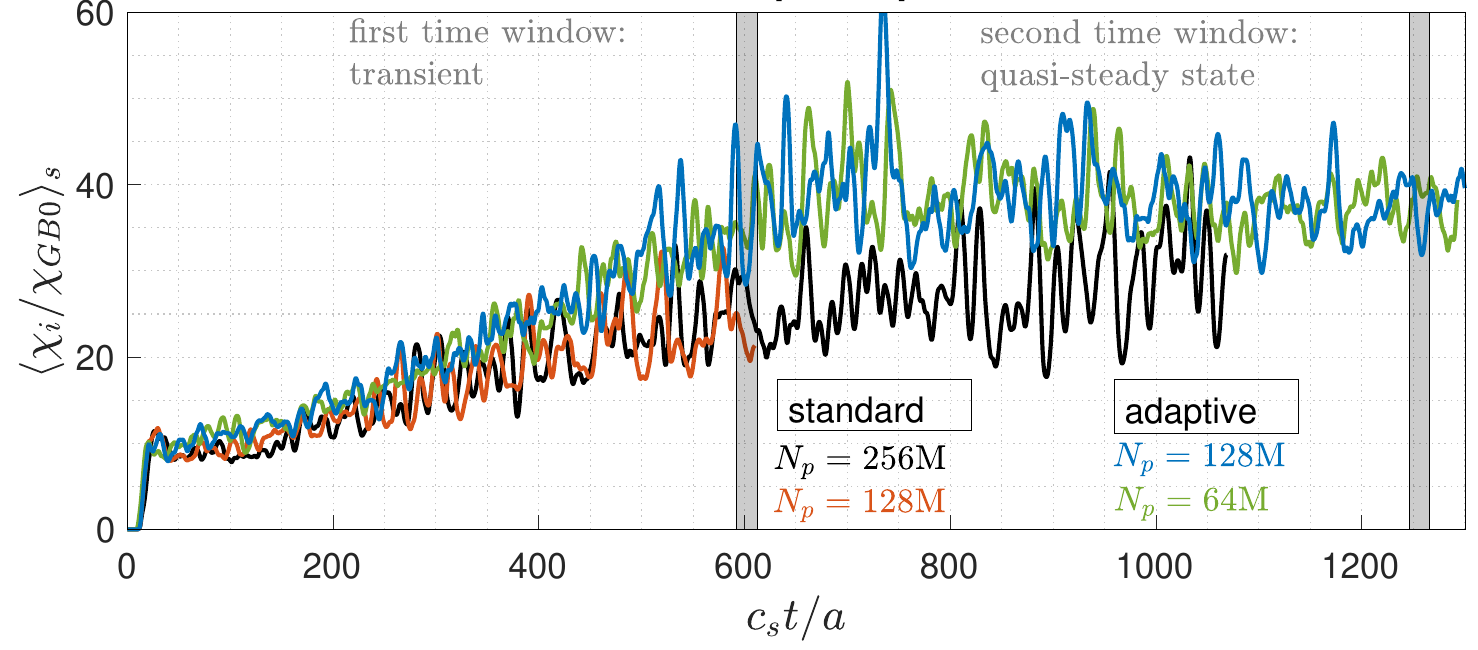}
	\end{subfigure}
	\begin{subfigure}[c]{1.0\columnwidth}
	    \caption{\label{fig:chiichie}}
	    \centering
		\includegraphics[width=1.0\linewidth]{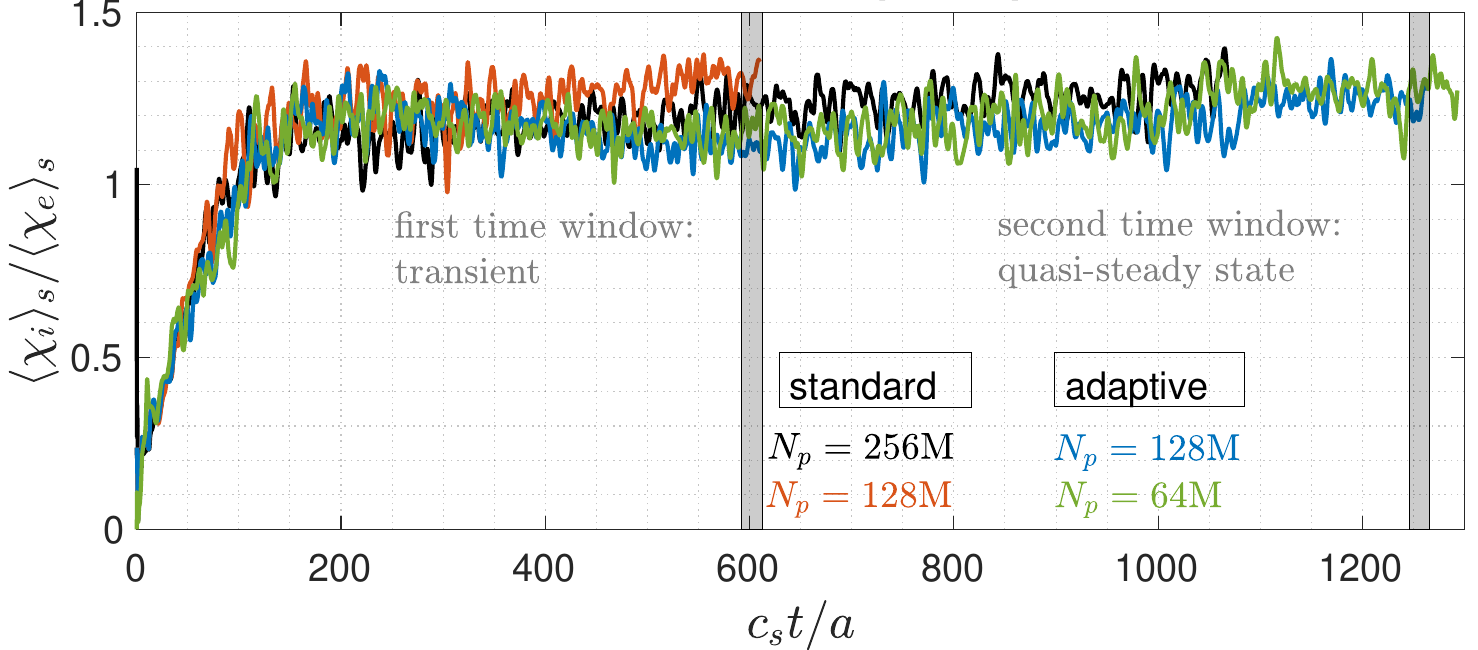}
	\end{subfigure}
	\caption{Radially averaged $s\in[0.7,0.9]$ (a) ion heat diffusivity and (b) ion-electron heat diffusivity ratio, for various marker numbers $N_p$ under the standard and adaptive schemes. Gray shaded areas represent two time windows for profile analysis.}
	\label{fig:chi}
\end{figure}

\begin{figure}[H]
    \centering
    \includegraphics[width=1.0\linewidth]{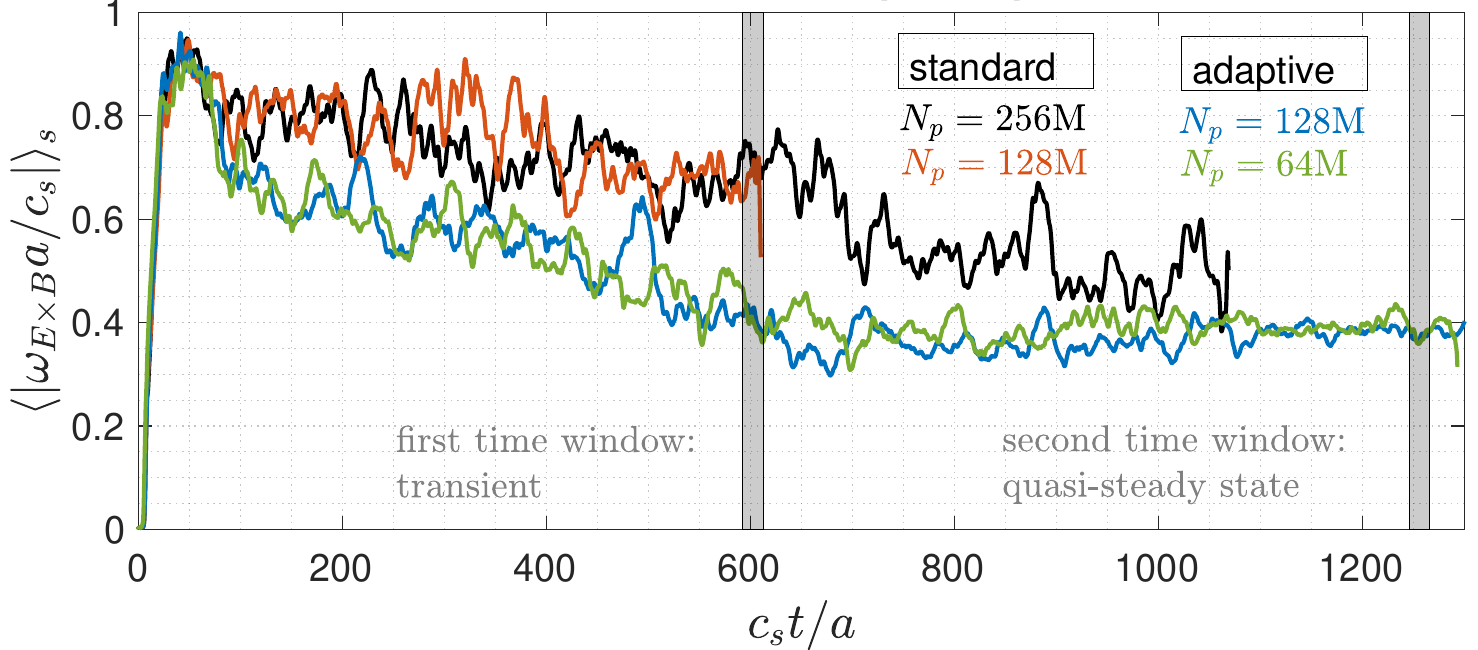}
    \caption{Absolute value of the $E\times B$ zonal flow shearing rate $\omega_{E\times B}$ radially averaged over $s\in[0.7,0.9]$ for various marker numbers $N_p$ under the standard and adaptive schemes. Gray shaded areas represent two time windows for profile analysis.}
    \label{fig:omega0d}
\end{figure}

The heat effective diffusivity $\chi$ is then given by
\begin{eqnarray}
    \chi &=& -\frac{\fsa{\vec{q}_{\rm H}\cdot\nabla\psi}}{n\partial T/\partial\psi\fsa{|\nabla\psi|^2}}, \nonumber
\end{eqnarray}
where $n$ and $T$ are the gyrocenter density and temperature respectively. The diffusivity is usually given in gyroBohm units $\chi_{GB0}=\rho_s^2c_s/a$ evaluated at $s=s_0=1$. Figure~\ref{fig:chi} shows the ion heat diffusivity $\chi_i$ radially averaged over $s\in[0.7,0.9]$ for all cases. Within the standard (non-adaptive) and adaptive cases, results appear converged in marker number $N_p$. The ion heat $\chi_i$ values of the adaptive cases are shown to be consistently higher than the standard cases. This difference between results of the two schemes can be explained by considering the shearing rate. Figure~\ref{fig:omega0d} shows the $s\in[0.7,0.9]$ radially averaged absolute value of the turbulence driven $E\times B$ zonal flow shearing rate, which is defined by~\cite{Hahm1994}
\begin{eqnarray}
    \omega_{E\times B}(s) &=& \frac{s}{2\psi_{\rm edge}q(s)}\pp{}{s}\left(\frac{1}{s}\pp{\fsa{\phi}}{s}\right). \nonumber
\end{eqnarray}

We see that the shearing rate for the standard cases are consistently higher than that of the adaptive cases. As $\omega_{E\times B}$ amplitudes measure the rate of turbulent eddy shearing, higher $\omega_{E\times B}$ amplitudes in the standard cases suppress turbulence more efficiently and thus allows for steeper gradients to be maintained for a given flux level, than in the adaptive cases within $s\in[0,7,0.9]$. (Notwithstanding the fact that in this work, no standard cases reached quasi-steady state, we assume that both standard and adaptive cases would share the same quasi-steady state, and thus the same average heat flux.) This explains the consistent higher $\chi_i$ of the adaptive cases as compared to the standard cases. As the system finally reaches quasi-steady state for $c_st/a\gtrsim1000$, simulations under both standard and adaptive schemes show signs of asymptoting to the same $\chi_i$ and $|\omega_{E\times B}|$ values. This asymptoting behaviour of the standard case is not conclusive as simulations with $N_p>256$M need to be run for an even longer simulation time in order to reach quasi-steady state for comparison with the adaptive cases presented here. Due to the numerical cost of such simulations, these standard runs are not conducted during the work of this paper.

To see how electron diffusivity $\chi_e$ compares to $\chi_i$, Fig.\ref{fig:chiichie} shows the ratio $\chi_i/\chi_e$ for all cases considered. In the time interval $c_st/a\in[0,150]$, the system evolves gradually from an initially TEM-dominated regime to a mixed TEM-ITG regime, with a $\chi_i/\chi_e$ ratio evolving from about $0.2$ to $1.2$. For further times, this ratio remains at the same value. This change of regime is due essentially to the density profile evolution, as will be discussed further (see Fig.\ref{fig:dens_short}).

\begin{figure}[H]
    \centering
    \includegraphics[width=1.0\linewidth]{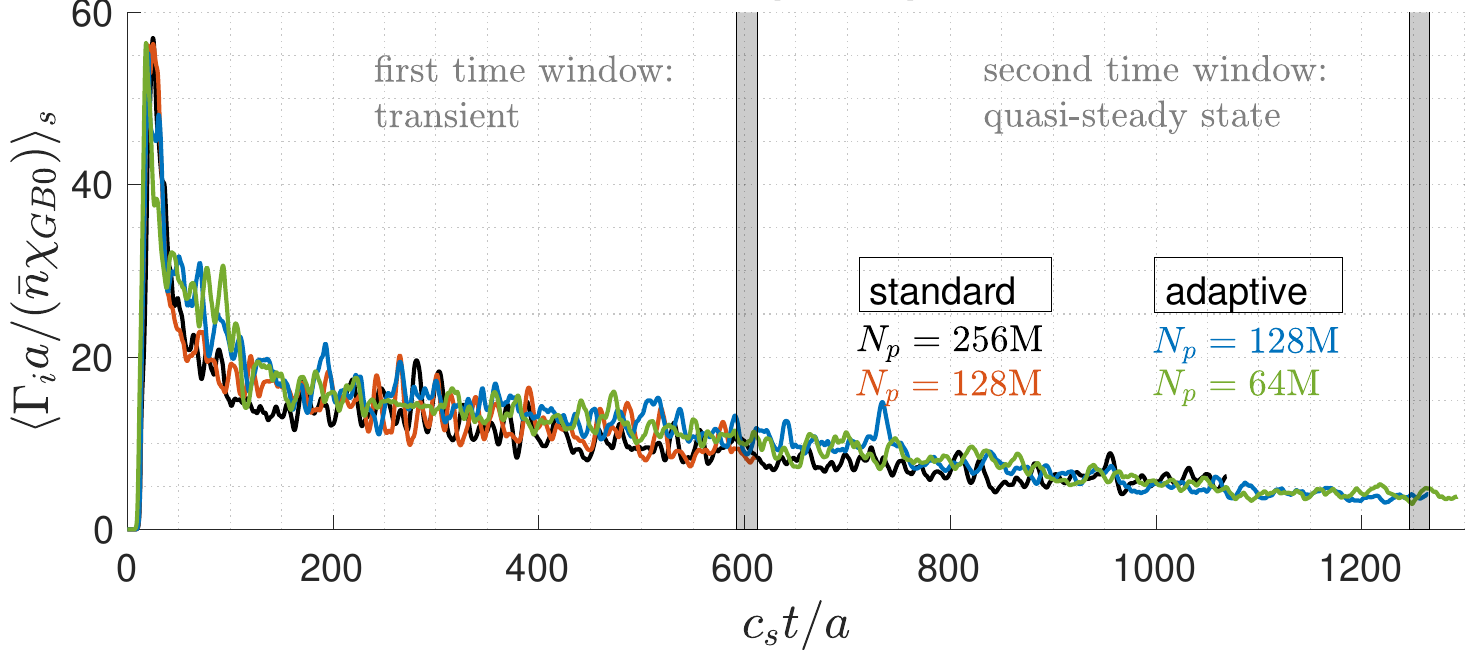}
    \caption{Ion gyrocenter flux radially averaged over $s\in[0.7,0.9]$ for various marker numbers $N_p$ under the standard and adaptive schemes. Gray shaded areas represent two time windows for profile analysis.}
    \label{fig:gammai}
\end{figure}

\begin{figure*}
	\centering
	\begin{subfigure}[c]{0.24\textwidth}
	    \caption{Standard, ions \label{fig:wrmsi_std128}}
		\centering
		\includegraphics[width=1.0\linewidth]{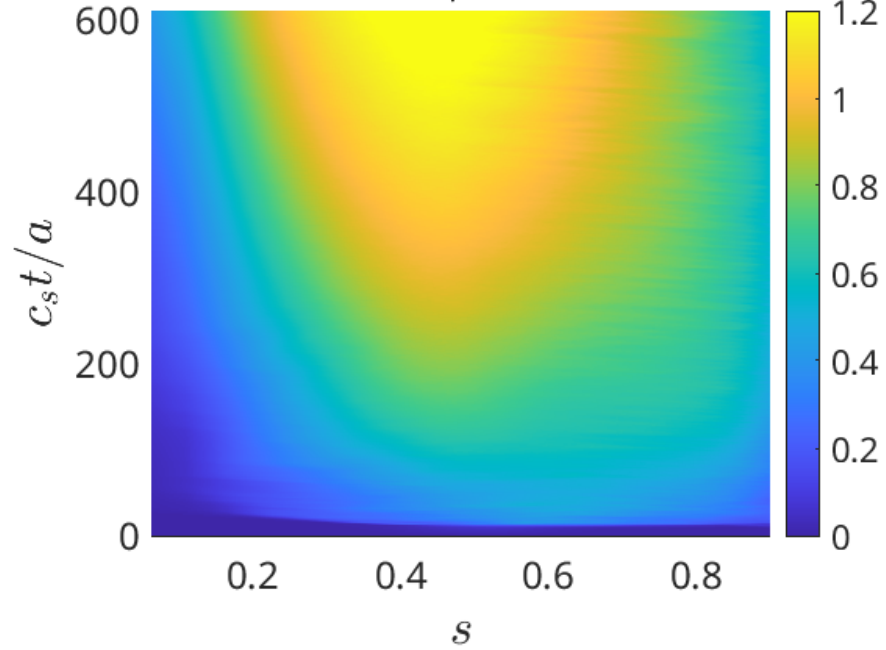}
	\end{subfigure}
	\begin{subfigure}[c]{0.24\textwidth}
	    \caption{Standard, electrons \label{fig:wrmse_std128}}
		\centering
		\includegraphics[width=1.0\linewidth]{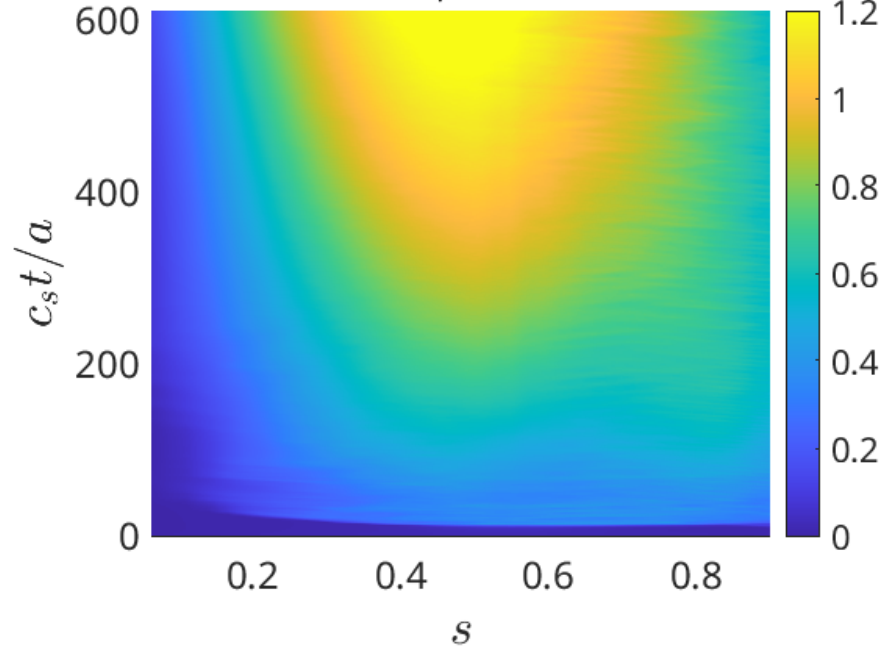}
	\end{subfigure}
        \begin{subfigure}[c]{0.24\textwidth}
	    \caption{Adaptive, ions \label{fig:wrmsi_adp128}}
		\centering
		\includegraphics[width=1.0\linewidth]{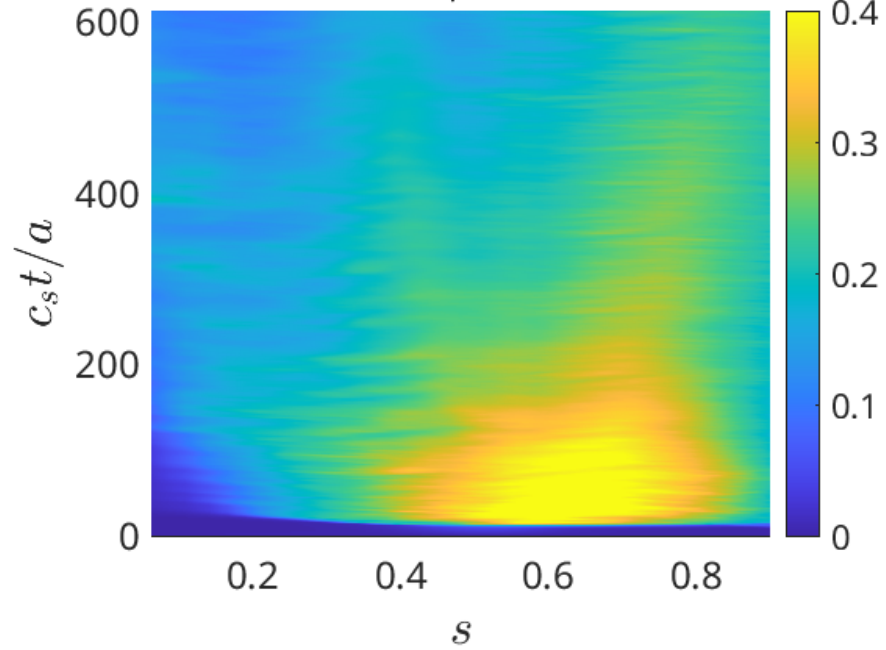}
	\end{subfigure}
	\begin{subfigure}[c]{0.24\textwidth}
	    \caption{Adaptive, electrons \label{fig:wrmse_adp128}}
		\centering
		\includegraphics[width=1.0\linewidth]{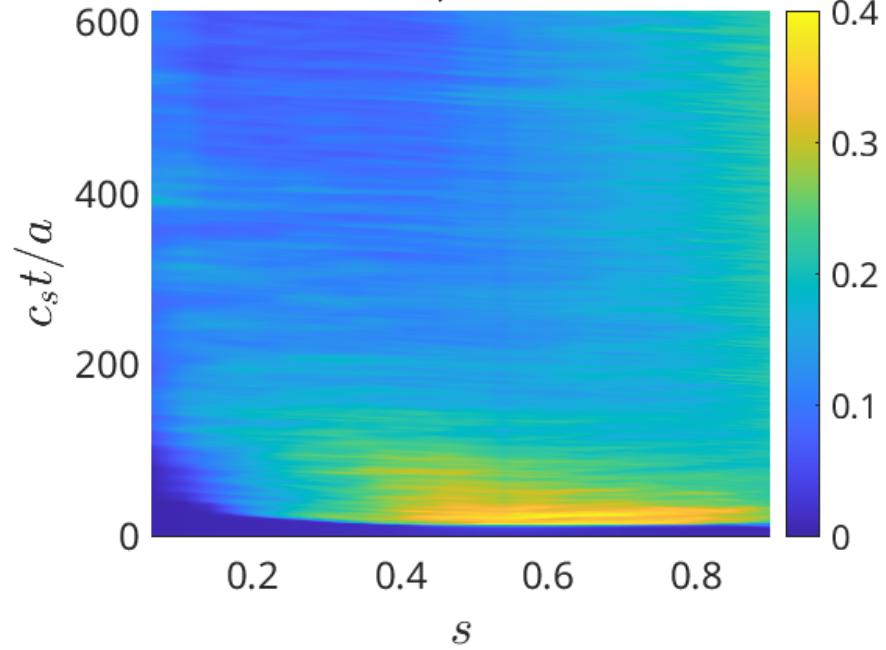}
	\end{subfigure}
	\caption{F.s.a.~weight standard deviation profiles $\sigma_w(s)=\sqrt{\fsa{w^2}(s)-\fsa{w}^2(s)}$ for marker number $N_p=128$M, under both the standard and adaptive schemes. Note the different colorbar scales between the standard and adaptive cases.}
	\label{fig:wrms}
\end{figure*}

As the system approaches quasi-steady state, with the buffer being the only source/sink of particles, the particle flux is expected to reduce and reach near zero values. Figure~\ref{fig:gammai} shows the f.s.a.~ion gyrocenter flux $\fsa{\vec{\Gamma}\cdot\nabla\psi}/\fsa{|\nabla\psi|}$ for cases with various $N_p$ under standard and adaptive schemes. We first note that adaptive cases (blue and green curves) have consistent slightly higher ion gyrocenter flux compared to the standard cases (black and orange) right after the initial overshoot in the time window $c_st/a\in[50,200]$. This is again consistent with the lower $\omega_{E\times B}$ amplitudes of the adaptive cases (see Fig.\ref{fig:omega0d}), thereby allowing for higher turbulence levels, and therefore, particle transport to develop. Nonetheless, the standard case with $N_p=256$M (black line) seems to initially converge to the same quasi-steady state value as that of the adaptive cases.

In ORB5, as in any $\delta f$-PIC code, the perturbed component of the distribution function for each species is expressed by the Klimontovich distribution
\begin{eqnarray}
    \delta f(\vec{Z},t) &=& \frac{N_{\rm ph}}{N_p}\sum_{p=1}^{N_p} \frac{w_p(t)}{2\pi\bstar{p}} \times \NL
    & & \delta[\vec{R}-\vec{R}_p(t)]\delta[\vp{}-\vp{p}(t)]\delta[\mu-\mu_p(t)], \label{eq:Klimontovich}
\end{eqnarray}
with $N_{\rm ph}=\int\dint{^3R}\,n_0(s)$ the total number of physical particles. The $p^{\rm th}$ marker phase space coordinates (with subscript $p$) follow the equations of motion (\ref{eq:eom}). The $w$-weight of the $p^{\rm th}$ marker is defined by $w_p(t)=\delta f(\vec{Z}_p,t)\Omega_pN_p/N_{\rm ph}$. Here, $\Omega_p$ is the phase-space volume associated to the $p^{\rm th}$ marker. The effectiveness of the adaptive control variate can be measured by estimating the standard deviation of the marker $w$-weights. The standard deviation indeed provides a measure of the ratio $\|\delta f\|/\|f\|$, which should remain low for the $\delta f$-PIC scheme to be statistically advantageous over a full-PIC approach. As diagnostic, we thus calculate the standard deviation of the weights in different radial bins. Given a radial grid $\{s\}$, the $i^{\rm th}$ bin is defined as $s_i\le s<s_{i+1}$. Let $s$ be the radial bin center in the following definition. We then define the local weight standard deviation as
\begin{eqnarray}
    \sigma_w(s) &=& \sqrt{\fsa{w^2}(s)-\fsa{w}^2(s)},\nonumber
\end{eqnarray}
with $\fsa{w}(s)=\sum_{p\in{\rm bin}}w_p/N_{\rm bin}$ and $\fsa{w^2}(s)=\sum_{p\in{\rm bin}}w_p^2/N_{\rm bin}$ the expectation value of $w$ and $w^2$ within that bin, respectively. Here, $N_{\rm bin}$ is the number of markers belonging to the bin. Note that, one expects convergence of $\sigma_w$ with marker number $N_p$ at a given time $t$. Furthermore, with dissipation in the form of the Krook operator $S_n$, the growth of $\sigma_w$ will be limited at quasi-steady state. Figure~\ref{fig:wrms} shows $\sigma_w$ for both standard and adaptive cases with marker number $N_p=128$M, up to a simulation time of $c_st/a=600$. From Figs.\ref{fig:wrmsi_std128} and \ref{fig:wrmse_std128}, we see that the radial location of increasing maximum values of $\sigma_w$ for both the ions and electrons are located where the profile deviation is the greatest (see Figs.\ref{fig:temp_short} and \ref{fig:dens_short}). After an initial phase where $\sigma_w$ grows (however to significantly lower values than in standard cases) the corresponding results under the adaptive scheme, on the other hand, have diminishing maxima. This is due to the adaptive control variate $f_0(t)$, with adaptive density and temperature profiles. Furthermore for the adaptive cases, we see a drop in electron $\sigma_w$ around $c_st/a=50$, compared to that of the ion which drops gradually after $c_st/a\simeq200$, where $\sigma_w$ is gradually decreased. This can be explained by considering Fig.\ref{fig:chiichie}, where we see that for $c_st/a\in[0,50]$, the electron diffusivity $\chi_e$ is more than $200\%$ that of the ion $\chi_i$. This suggests a quicker evolving electron temperature profile $T_e$ during this period. Thus, as the adaptation rate is sufficiently high for this case, the electron $\sigma_w$ is reduced at a shorter time scale. Comparing results between the standard and adaptive cases, we see that the latter only have $15\%$ the maximum value of $\sigma_w$ compared to that of the former. Furthermore, at $c_st/a=600$, $\sigma_w$ for the standard cases is still growing, while that of the adaptive cases exhibit a steady value of around $0.2$.

\begin{figure}[H]
    \centering
    \includegraphics[width=1.0\linewidth]{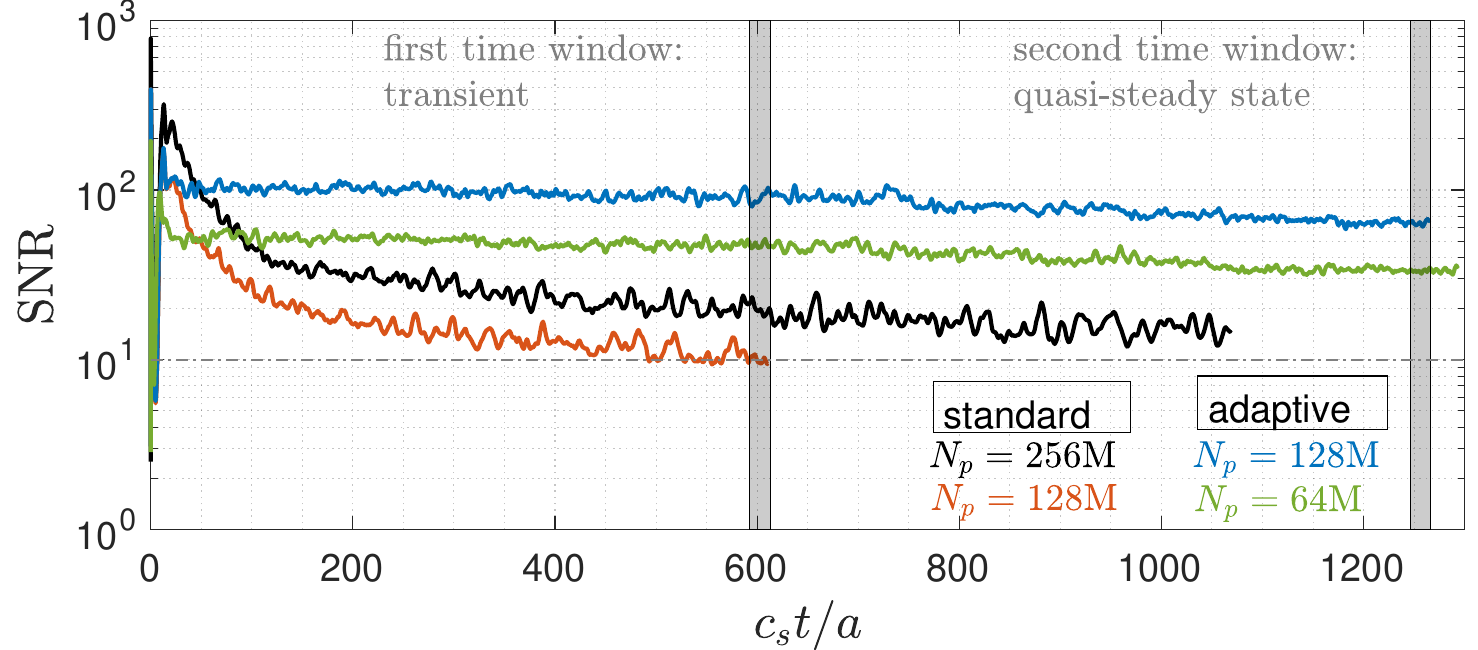}
    \caption{Time dependence of global Signal-to-Noise Ratio (SNR) values, for various marker numbers $N_p$ under the standard and adaptive schemes. The signal includes the zonal component $(m,n)=(0,0)$. Corresponding values without the zonal component are about $10\%$ lower for all cases and for all times. Horizontal dashed line indicates the empirically set minimum value of $10$ for ensuring quality simulations. Gray shaded areas represent two time windows for profile analysis.}
    \label{fig:snr}
\end{figure}

\begin{figure*}
	\centering
	\begin{subfigure}[c]{0.24\textwidth}
	    \caption{Standard\newline \label{fig:omega2D_std128_rhs}}
	    \centering
		\includegraphics[width=1.0\linewidth]{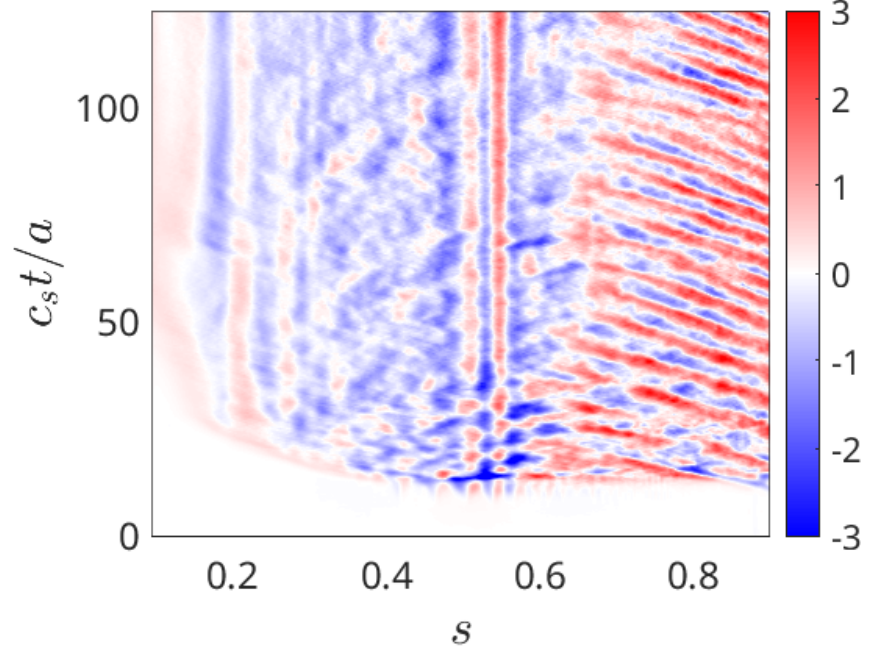}
	\end{subfigure}
	\begin{subfigure}[c]{0.24\textwidth}
	    \caption{Adaptive, with QNE r.h.s.~correction\label{fig:omega2D_adp128_analytic}}
	    \centering
		\includegraphics[width=1.0\linewidth]{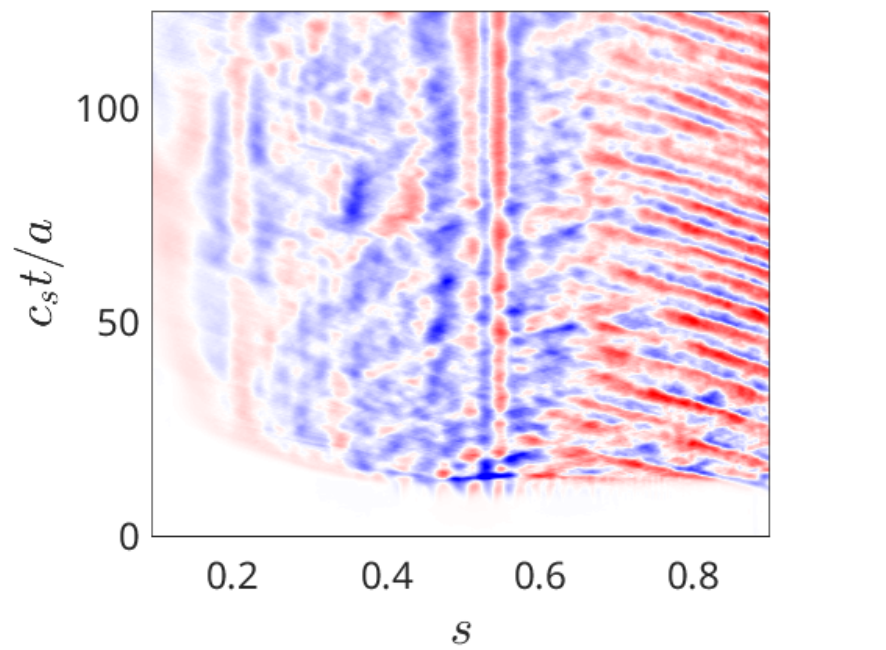}
	\end{subfigure}
    \begin{subfigure}[c]{0.24\textwidth}
	    \caption{Adaptive, coupled background electron density\label{fig:omega2D_adp128_fsa}}
	    \centering
		\includegraphics[width=1.0\linewidth]{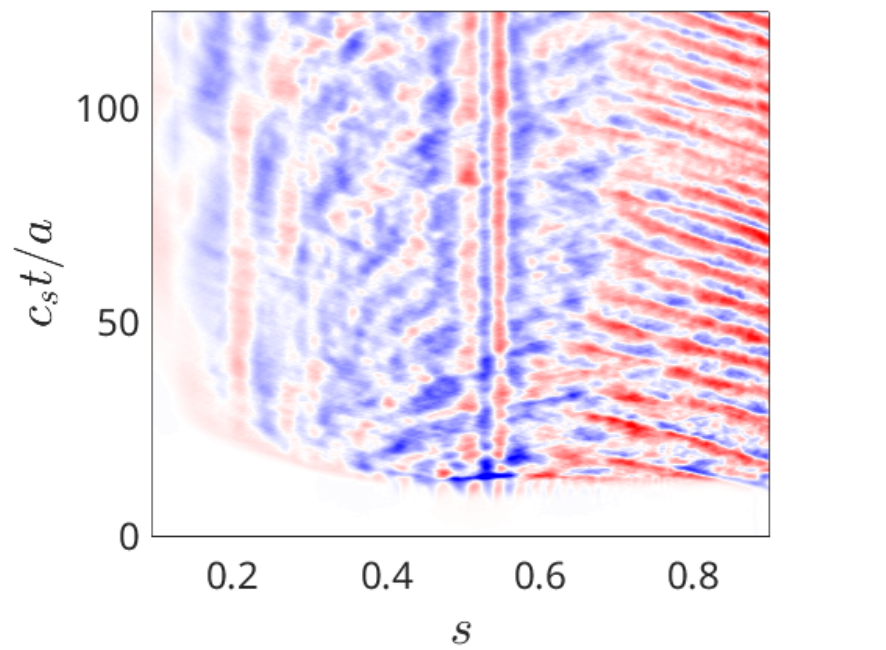}
	\end{subfigure}
    \begin{subfigure}[c]{0.24\textwidth}
	    \caption{Adaptive, without QNE r.h.s.~correction\label{fig:omega2D_adp128_xrhs}}
	    \centering
		\includegraphics[width=1.0\linewidth]{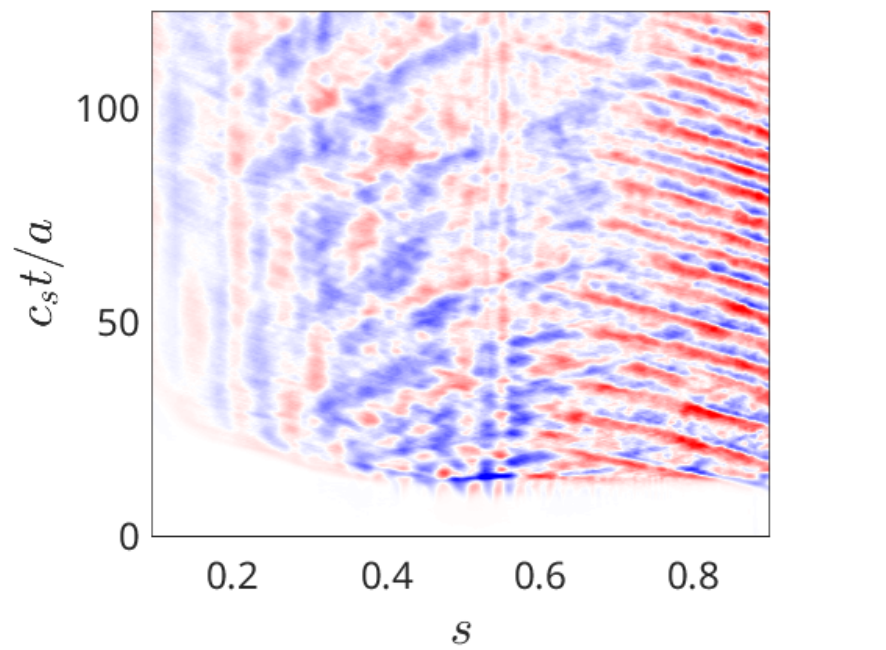}
	\end{subfigure}
	\caption{Evolution up to time $c_st/a=123$ of the radial profile of the zonal flow $E\times B$ shearing rate $\omega_{E\times B}$ with marker number $N_p=128$M. (a) Standard (non-adaptive). Adaptive cases (b) including correction term (\ref{eq:qne_corr}) calculated via quadratures, (c) imposing Eq.(\ref{eq:qnrhs_fsa}) and (d) neglecting correction term but still evolving $f_{0i}$ and $f_{0e}$ without imposing Eq.(\ref{eq:qnrhs_fsa}). All plots share the same color scale.}
	\label{fig:qnrhs}
\end{figure*}

Next, we consider the Signal-to-Noise Ratio (SNR) values. This diagnostic is based on a Fourier filter on the $(m,n)$-spectrum of the r.h.s.~of Eq.(\ref{eq:qne_hyb}) by considering the gyrokinetic assumption $|k_\parallel/k_\perp|\ll1$, with $k_\parallel$ and $k_\perp$ the amplitudes of the parallel and perpendicular components of the perturbation wave vector to the equilibrium magnetic field $\vec{B}$, respectively. The method of calculation is explained in Ref.~\cite{Murugappan2022}. Figure~\ref{fig:snr} shows the SNR time trace for all cases discussed. The SNR values of the standard cases with $N_p=256$M (black) and $N_p=128$M (orange) can be seen to fall continuously with similar rates as simulation time passes. The gain in SNR value achieved by increasing $N_p$ is proportional to $N_p$, owing to the fact that this diagnostic is based on squared fluctuation amplitudes and in particular the noise estimate scales as $1/N_p$. Due to degrading simulation quality, simulations are stopped when SNR values reach the empirically set minimum threshold~\cite{Bottino2007,Jolliet2009,Jolliet2012} of $10$. The trends displayed in Fig.~\ref{fig:snr} imply that in order to reach a similar SNR as the one at the end of the adapted case with $N_p=64$M, one would need at least $N_p=512$M when using the non-adaptive scheme. Thus the adaptive scheme allows for a reduction in computational cost by a factor of $8$ for ensuring the same numerical quality. For a given $N_p$, the adaptive cases have their SNR values drop much less from their maximum values compared to the standard cases. This drop happens mostly at the initial phase $c_st/a\lesssim200$ of the simulation when profiles evolve the most. This reduction could be lessened somewhat further by increasing the adaptation rates $\alpha_n$ and $\alpha_E$ of Eqs.(\ref{eq:relax_dens}) and (\ref{eq:relax_ekin}), though improvements resulting from increasing adaptation rates have been found to be marginal. This is because the adaptive scheme of this work is based on a f.s.a.~control variate. Any variation of the fluctuations in the poloidal direction, for example, will not be accounted for in $f_0$. Nonetheless, for the cases studied, we conclude that a simulation run with $N_p=64$M, or even potentially $N_p=32$M, gives us significant results, which otherwise, i.e.~with standard scheme, would only be obtained with at least $N_p=512$M.

Finally, the importance of the correction term given by Eq.(\ref{eq:qne_corr}) to the quasi-neutrality equation (\ref{eq:qne_hyb}) is illustrated by its effect on the zonal flow shearing rate $\omega_{E\times B}$. Figure~\ref{fig:qnrhs} shows the time evolution up to $c_st/a=123$ of the radial profile of $\omega_{E\times B}(s)$ for four simulations run with marker number $N_p=128$M. As the plots shown are still early in simulation times, according to Fig.\ref{fig:snr}, they are considered reliable, whether under the standard or adaptive schemes. Using the standard case in Fig.\ref{fig:omega2D_std128_rhs} as reference, we first note that strong inward avalanches are triggered from the edge, in a region where the time-averaged $\omega_{E\times B}$ has positive values. The inward direction of avalanche propagation is in line with the analysis presented in Ref.~\cite{McMillan2009}. We also observe a stationary radial corrugation structure coinciding with the $q=1$ flux surface at $s=0.538$. Small transport barriers then to develop around low order Mode-Rational-Surfaces (MRS) including corrugated zonal flow shearing rate profiles. This effect, related to the non-adiabatic passing electron dynamics, is only partially captured by our hybrid electron model, given that it only accounts for the f.s.a.~kinetic contribution from passing particles (see Eq.(\ref{eq:qne_hyb}). Such structures have been previously analyzed~\cite{Dominski2017} using either a Pad\'{e} or an arbitrary wavelength solver for the ion polarisation density, and using fully drift kinetic electrons. The role of the correction term (\ref{eq:qne_corr}) is shown by comparing Figs.\ref{fig:omega2D_adp128_analytic}-\ref{fig:omega2D_adp128_xrhs} against the reference case in Fig.\ref{fig:omega2D_std128_rhs}. As the control variate is close to a f.s.a.~function, one expects persistent f.s.a.~features to be gradually erased as low order f.s.a.~velocity moments are being removed from the $\delta f$ component, which appears on the r.h.s.~of Eq.(\ref{eq:qne_hyb}). This explains that without the correction term, Fig.\ref{fig:omega2D_adp128_xrhs}, the simulation fails to resolve these persistent corrugated structures. Based on Fig.\ref{fig:omega2D_adp128_fsa}, the prescription of the adapted electron density given by Eq.(\ref{eq:qnrhs_fsa}) indeed preserves the persistent f.s.a.~structures. Though not shown, adaptive runs with QNE correction term calculated by quadratures (Fig.\ref{fig:omega2D_adp128_analytic}) or by means of Eq.(\ref{eq:qnrhs_fsa}) are indistinguishable under the simulation parameters used in this work.

\subsection{Evolved profiles comparison} \label{sec:early}

\begin{figure}[H]
	\centering
	\begin{subfigure}[c]{0.9\columnwidth}
	    \caption{Ions \label{fig:RLTi_short}}
	    \centering
		\includegraphics[width=1.0\linewidth]{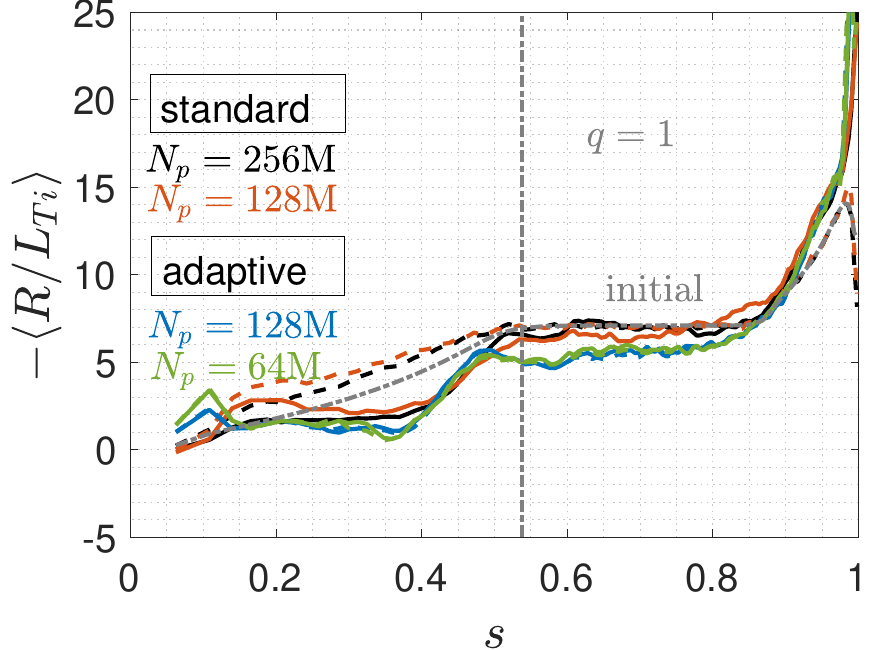}
	\end{subfigure}
	\begin{subfigure}[c]{0.9\columnwidth}
	    \caption{Electrons \label{fig:RLTe_short}}
	    \centering
		\includegraphics[width=1.0\linewidth]{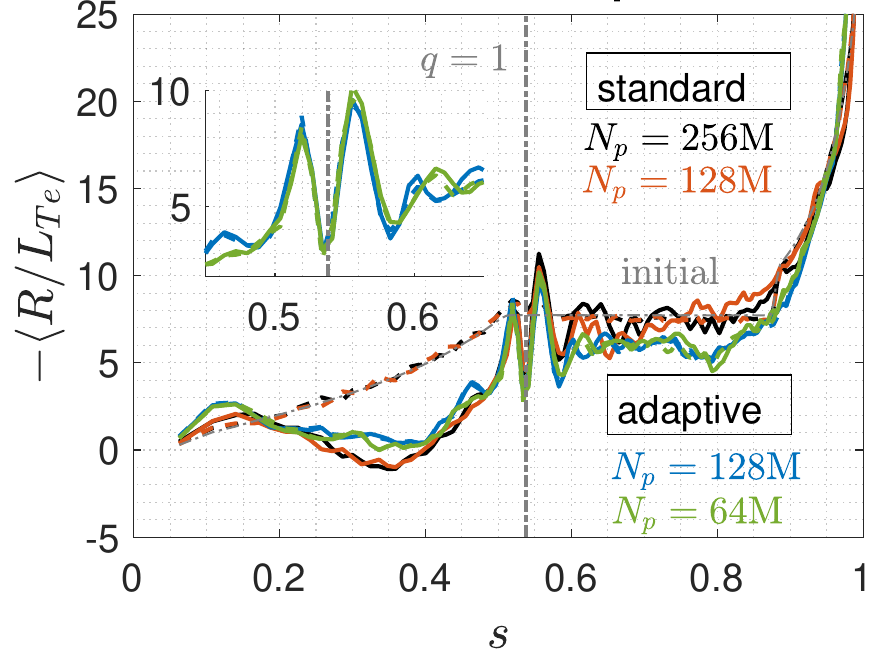}
	\end{subfigure}
	\caption{Time averaged $c_st/a\in[592,613]$ f.s.a.~gyrocenter temperature logarithmic gradient profiles for various marker numbers $N_p$ under the standard and adaptive schemes for (a) ions and (b) electrons. Dashed and solid curves show contribution by $f_0$ and $f_0+\delta f$ respectively. Initial profiles given by dotted gray curve. Vertical dashed line indicates position of $q=1$ surface.}
	\label{fig:rlt_short}
\end{figure}

\begin{figure}[H]
	\centering
	\begin{subfigure}[c]{0.49\columnwidth}
	    \caption{Ions \label{fig:tempi_short}}
	    \centering
		\includegraphics[width=1.0\linewidth]{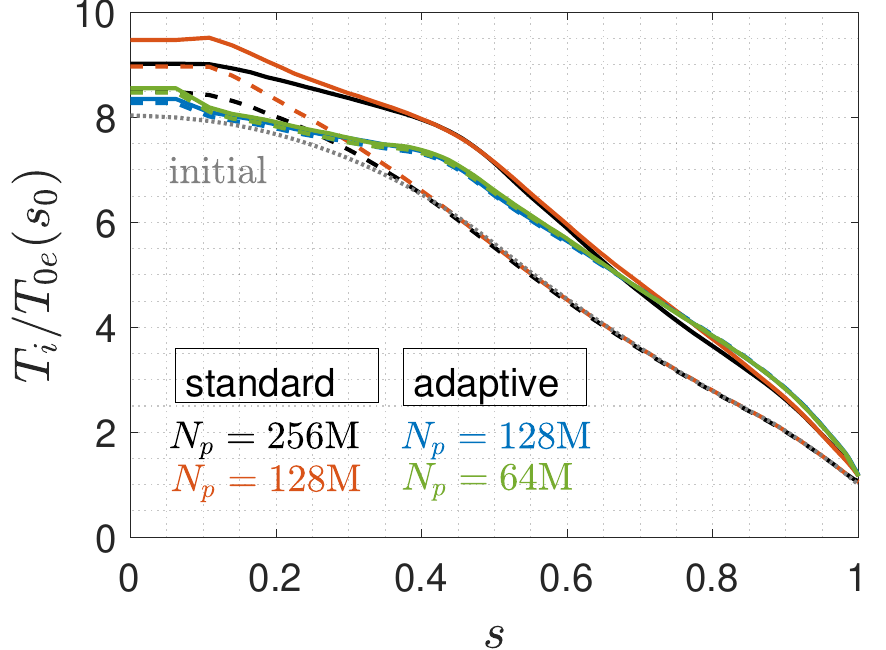}
	\end{subfigure}
	\begin{subfigure}[c]{0.49\columnwidth}
	    \caption{Electrons \label{fig:tempe_short}}
	    \centering
		\includegraphics[width=1.0\linewidth]{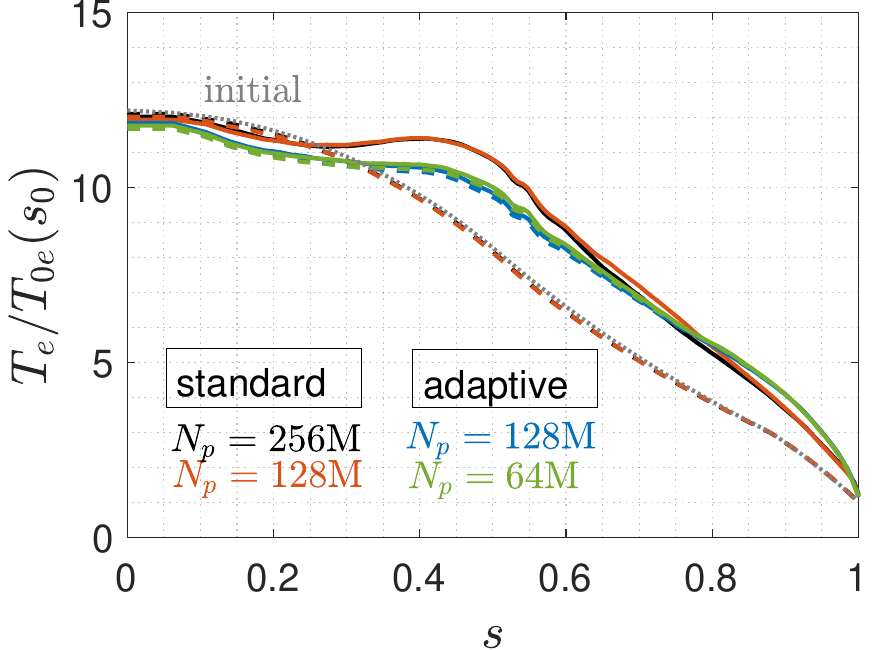}
	\end{subfigure}
	\caption{F.s.a.~gyrocenter temperature profiles time averaged over $c_st/a\in[592,613]$ for various marker numbers $N_p$ under the standard and adaptive schemes. Dashed and solid lines show contribution by $f_0$ and $f_0+\delta f$ respectively, essentially on top of each other in adaptive cases.}
	\label{fig:temp_short}
\end{figure}

We now consider evolved f.s.a.~profiles, further averaged over the first time window $c_st/a\in[592,613]$ (shown e.g.~in Fig.~\ref{fig:chi}). All profiles are derived from binning marker $f_0\Omega$ values and $w$-weights into radial bins for the contributions from the background $f_0$ and perturbed $\delta f$ distributions respectively. Figure~\ref{fig:rlt_short} shows the gyrocenter temperature logarithmic gradient for all cases considered and for both ions and electrons. We first note the dip towards zero in logarithmic gradients for both species inwards from the heat sources (see Fig.~\ref{fig:source}) at $s=0.35$. This dip moves further towards the magnetic axis at later times as the heat is transported from the heat source peak towards the core, thus flattening the temperature profiles in this region (also see Fig.\ref{fig:temp_short}). Figure~\ref{fig:RLTi_short} shows a steepening of gradient in $s\in[0.9,1.0]$ (well within the buffer region) for the ions, which is barely present for the electrons (in Fig.\ref{fig:RLTe_short}). On the other hand, in the adaptive cases Fig.~\ref{fig:RLTe_short} shows a corrugation of the electron logarithmic gradient in the vicinity of the $q=1$ flux surface at $s\approx0.55$. This can be further seen in the inserted figure in Fig.~\ref{fig:RLTe_short}, showing a zoom-in of this region and illustrating the ability of the evolving background to capture such fine profile features. No such corrugation is observed for the ions. This corrugated structure is related to the one of the $\omega_{E\times B}$ radial profiles of Fig.~\ref{fig:qnrhs}. Finally, we see that for both species, the adaptive cases resulted in a lower logarithmic gradient in  region $s\in[0.6,0.9]$. This is related to the fact that adaptive cases exhibit greater local heat diffusivity values (see Fig.\ref{fig:chii}) due to lower local $\omega_{E\times B}$ amplitudes, as mentioned.

Ion and electron temperature profiles are shown in Fig.\ref{fig:temp_short}. Note that while at this point in time, there is no off-axis peak ion temperature profile, there is one for the electron temperature under the standard scheme, at around $s=0.45$. This does not coincide with the peak of the heat source for the electrons at around $s=0.55$ (see Fig.\ref{fig:sourcee}). The slight corrugation for the electron temperature at $s=0.55$ was emphasized in its logarithmic gradient in Fig.~\ref{fig:RLTe_short}. One further notes an increase in ion temperature at the magnetic axis under the standard scheme, with the $N_p=128$M case having a larger increase than that of the $N_p=256$M simulation. Only a slight increase in ion temperature at the magnetic axis under the adaptive scheme is observed. We thus attribute this problem as numerical, and is related to the under-sampling of phase space volume discussed in Sec.~\ref{sec:pvol}. This is already suggested by the dip in background temperature profiles (dashed) of the standard cases (black and orange). As these background profiles are not adaptive, we expect them to coincide with the initial profile (dashed gray). Between the standard cases, the case with $N_p=128$M deviates more from its initial values compared to the case with $N_p=256$M. This effect is not observed for electrons. This sampling problem of phase-space volume will be discussed in Sec.\ref{sec:pvol}.

\begin{figure}[H]
	\centering
	\begin{subfigure}[c]{0.49\columnwidth}
	    \caption{\label{fig:densi_short}}
	    \centering
		\includegraphics[width=1.0\linewidth]{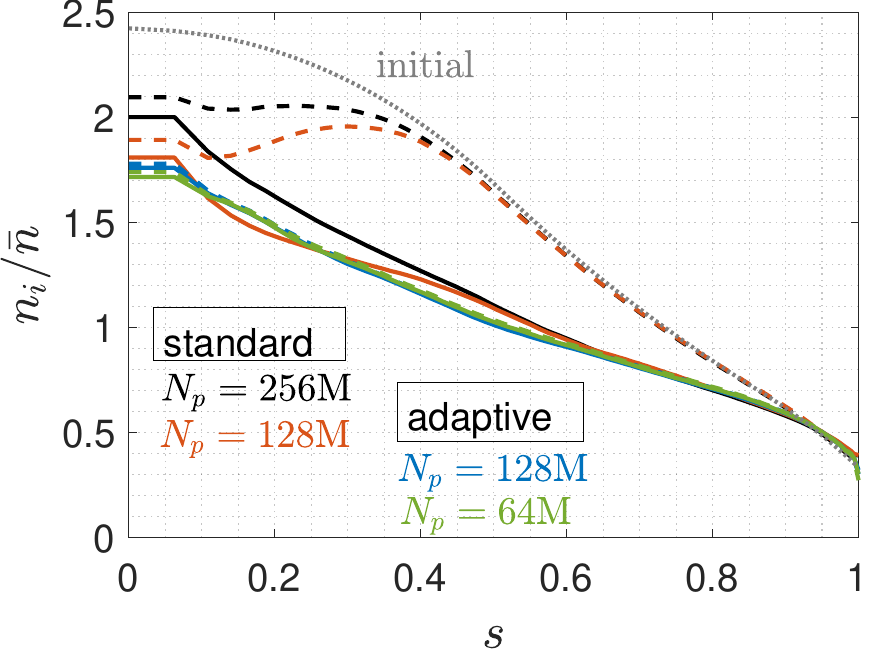}
	\end{subfigure}
	\begin{subfigure}[c]{0.49\columnwidth}
	    \caption{\label{fig:ddens_short}}
	    \centering
		\includegraphics[width=1.0\linewidth]{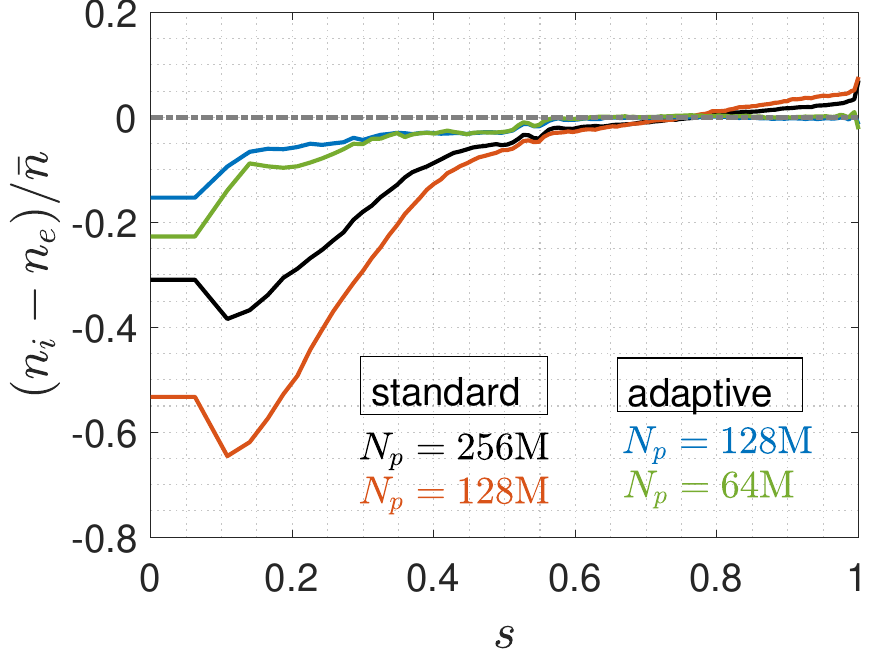}
	\end{subfigure}
	\caption{(a) ion gyrocenter density profiles and (b) ion-electron gyrocenter density difference, time averaged over $c_st/a\in[592,613]$ for various marker numbers $N_p$ under the standard and adaptive schemes. Dashed and solid lines show contribution by $f_0$ and $f_0+\delta f$ respectively.}
	\label{fig:dens_short}
\end{figure}

Figure \ref{fig:dens_short} shows the gyrocenter densities of the ions and electrons time-averaged over $c_st/a\in[592,613]$. Focusing first on Fig.~\ref{fig:densi_short}, we see that at the considered stage of the simulation, the non-adaptive background density of the standard cases (dashed black and orange) has a lower value near the magnetic axis as compared to the initial profile (dashed gray). The source of this deviation is the same as that in Fig.~\ref{fig:tempi_short} for the ion temperature, i.e.~under-sampling of phase space. Next, the $f_0$ contribution to density of the adaptive cases (dashed blue and green) is nearly identical to the one including the $\delta f$ contribution, which indicates that the time-dependent background density has correctly captured the evolution of the total f.s.a.~density. Furthermore, the adaptive cases seem to have converged in marker number $N_p$, whereas the standard cases has not, with $N_p=256$M for the standard case showing a smaller deviation of the evolved density profile compared to the $N_p=128$M case. Figure~\ref{fig:ddens_short} shows the difference between ion gyrocenter and electron densities, where we see a larger difference between these densities for the standard cases. Between these standard cases, the case with $N_p=128$M shows a larger density difference as compared to that with $N_p=256$M, thus indicating that this density difference is partly due to sampling error, as pointed out in Figs.~\ref{fig:temp_short} and \ref{fig:densi_short}. The same $N_p$ trend can be observed for the adaptive cases, though the results are clearly already more converged. Note that ion gyrocenter and electron density difference is not expected to be zero, as one still needs to account for ion Finite-Larmor-Radius (FLR) effects and ion polarisation density. Taking a step back, we now see in Fig.~\ref{fig:densi_short} that the apparent convergence of the standard case with $N_p=128$M (solid orange) with the adaptive cases (solid blue and green) is just a coincidence.

\subsection{Phase-space volume sampling} \label{sec:pvol}
\begin{figure*}
	\centering
	\begin{subfigure}[c]{0.24\textwidth}
	    \caption{Standard, $N_p=256$M \label{fig:dpvoli_std256_short}}
		\centering
		\includegraphics[width=1.0\linewidth]{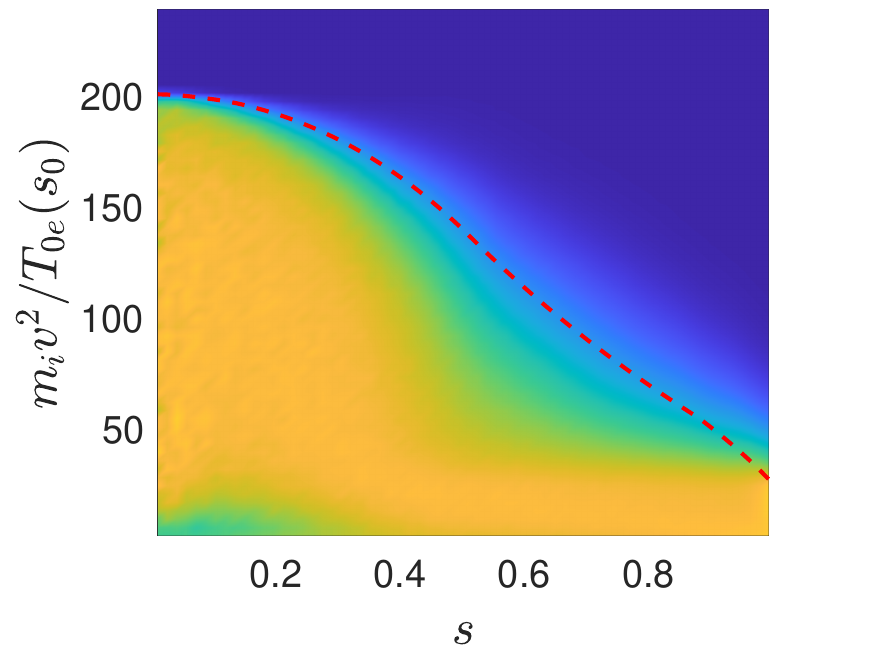}
	\end{subfigure}
	\begin{subfigure}[c]{0.24\textwidth}
	    \caption{Standard, $N_p=128$M \label{fig:dpvoli_std128_short}}
		\centering
		\includegraphics[width=1.0\linewidth]{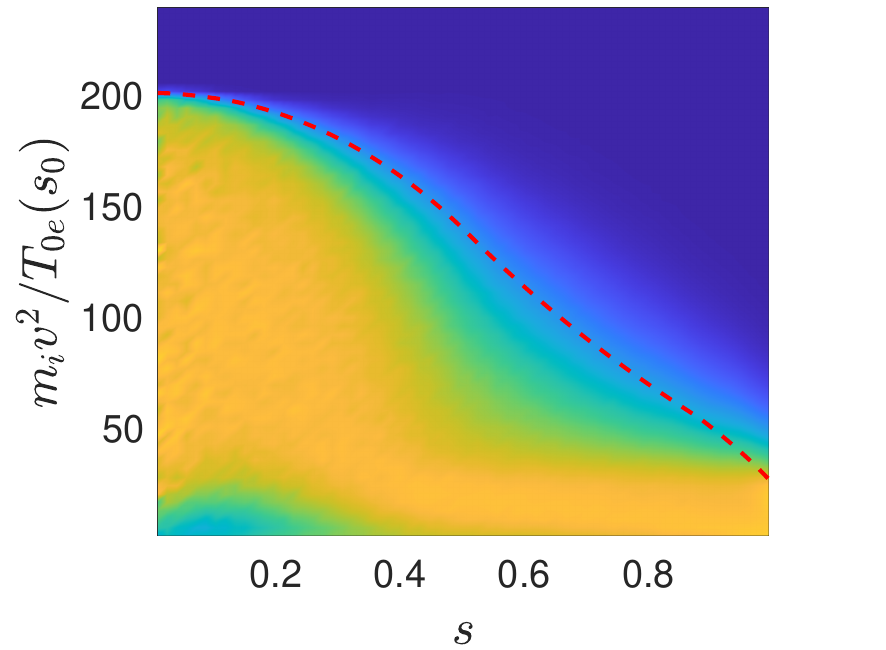}
	\end{subfigure}
	\begin{subfigure}[c]{0.24\textwidth}
	    \caption{Adaptive, $N_p=128$M \label{fig:dpvoli_adp128_short}}
		\centering
		\includegraphics[width=1.0\linewidth]{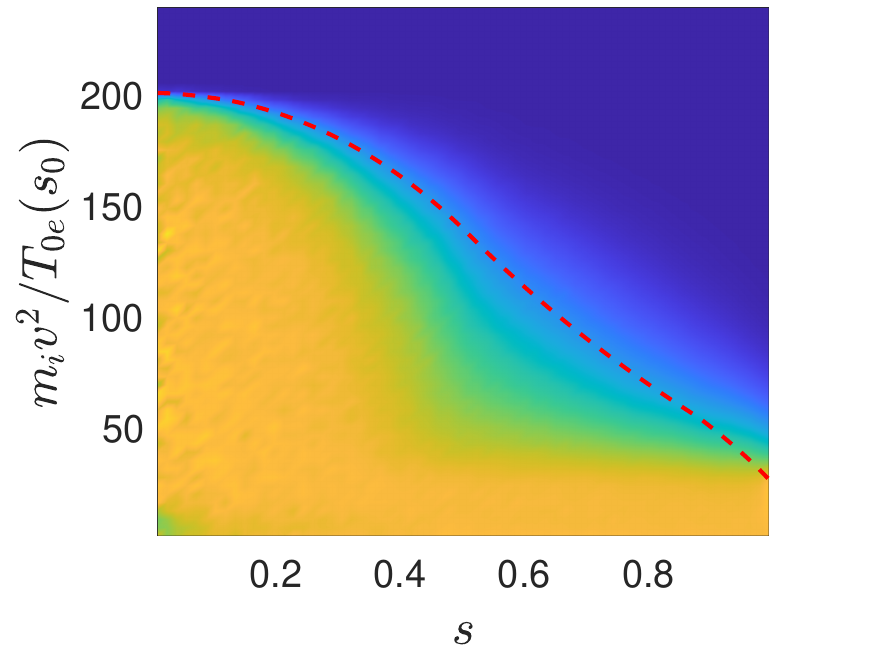}
	\end{subfigure}
	\begin{subfigure}[c]{0.24\textwidth}
	    \caption{Adaptive, $N_p=64$M \label{fig:dpvoli_adp64_short}}
		\centering
		\includegraphics[width=1.0\linewidth]{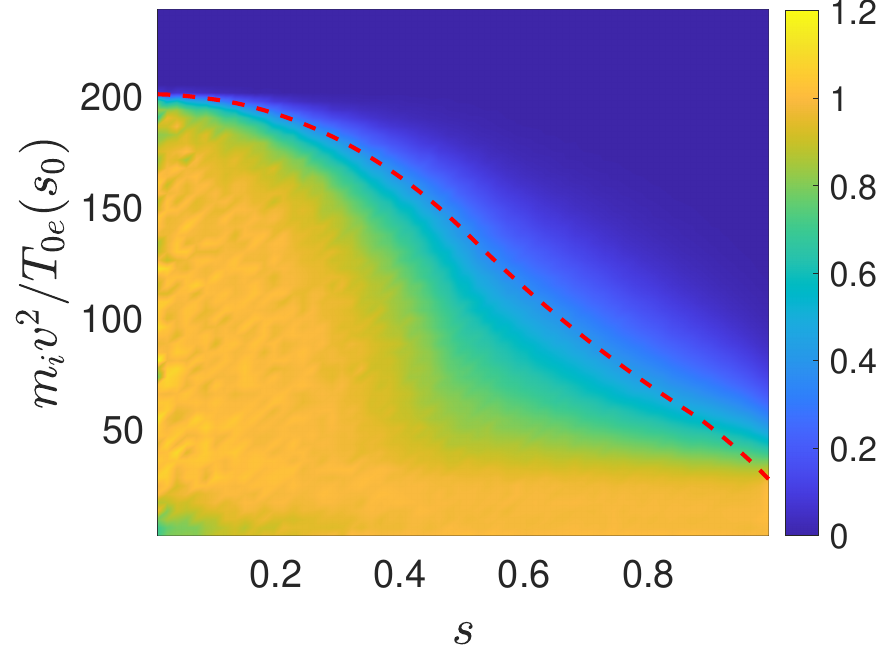}
	\end{subfigure}
	\caption{Ion phase-space volume diagnostic applied to Cartesian bins in $(s,v^2)$-space for the standard and adaptive cases for various marker numbers $N_p$ at $c_st/a=612$. Shown is the ratio $\Omega(t)/\Omega^{\rm (a)}$. Values near $1$ reflect good sampling, while ratios deviating significantly from $1$ reflect poor/deficient sampling. The red dashed line on the contour plots indicates the energy per mass upper-bound during initial marker loading, given by $\kappa_v^2v_{\rm th}^2(s)$. All plots share the same colour scale.}
	\label{fig:dpvoli}
\end{figure*}

The configuration space $\vec{R}=[s,\ts{},\varphi]$ volume enclosed by the magnetic surface labeled by $s$ be given by
\begin{eqnarray}
    V(s) &=& \int\dint{^3R} = \int_0^s\dint{s}\int_0^{2\pi}\dint{\ts{}}\int_0^{2\pi}\dint{\varphi}\,J_s(s,\ts{}), \label{eq:V}
\end{eqnarray}
where $\dint{^3R}=J_s\dint{s}\dint{\ts{}}\dint{\varphi}$ is the configuration space differential volume element, and prime denotes the derivative of its argument. At the beginning of each simulation, markers are distributed uniformly in $V(s=1)$. In the velocity $(\vp{},v_\perp)$-space, markers are distributed uniformly in a semi-circular disk of radius $\kappa_vv_{\rm th}(s)$, $\vp{}^2+v_\perp^2\le\kappa_v^2v_{\rm th}^2(s)$, with $v_\perp\ge0$. Here, $\kappa_v=5$ is set for all simulations of this work. As the simulation evolves over time, markers move in phase space according to the equations of motion (\ref{eq:eom}), and the overall marker distribution should behave like an incompressible fluid. Due to time-integration inaccuracies, markers will deviate from their exact orbit and as a result incompressibility is not exactly ensured, reflected by increasing errors in phase space sampling over time. This is particularly true near the magnetic axis $s=0$ where, according to the current loading, there are very few markers per $s$-bin. A useful diagnostic is to bin the phase-space volume $\Omega_p$ of each marker onto a grid of reduced $[s,v^2]$-phase-space. Let $(i,j)$ be the index of such a bin $\Gamma_{ij}=[s_i,s_{i+1}]\times[v_j^2,v_{j+1}^2]$. The exact phase space volume of the $(i,j)^{\rm th}$ bin is given by
\begin{eqnarray}
    \Omega_{ij}^{\rm (a)}) &=& \int_{\Gamma_{ij}}\dint{\Omega} = \frac{4\pi(v_{j+1}^3-v_j^3)}{3}\int_{s_i}^{s_{i+1}}\dint{s}\, V'(s), \NL
    & &\label{eq:pvol}
\end{eqnarray}
where $V'(s)$ is the derivative of the configuration space volume Eq.(\ref{eq:V}). Figure \ref{fig:dpvoli} shows the binned ion phase-space volume $\Omega_{ij}(t)=\sum_{p\in\Gamma ij}\Omega_p(t)$ normalized to the reference value Eq.(\ref{eq:pvol}) at time $c_st/a=613$. A feature common to all cases is that there is a diffusion of sampled phase-space volume $\Omega$ around the velocity cut-off at $\kappa_vv_{\rm th}$, as expected, with greatest diffusion around $s\in[0.5,0.6]$. Focusing on the standard cases, Figs.\ref{fig:dpvoli_std256_short}-\ref{fig:dpvoli_std128_short}, we see an under-sampling in $\Omega$ around $s\in[0,0.3]$ at low energies. We also see that the situation is somewhat improved with higher $N_p$, indicating that this problem is of numerical origin. The same is true for the adaptive cases presented in Figs.\ref{fig:dpvoli_adp128_short}-\ref{fig:dpvoli_adp64_short}, though the extent of the under-sampling is much smaller compared to the standard cases. To determine if this phase-space under-sampling is due to integration errors in marker trajectories or phase-space volume corruption, we consider now the marker distribution in $(s,v^2)$-space. The expression for the loaded marker number $\dint{N}$ in the differential area $\dint{s}\dint{v^2}$ at initial time is given by
\begin{eqnarray}
    \dint{N} &=& \dint{s}\dint{v^2}\,\frac{N_p}{V(s=1)}\frac{V'(s)}{\kappa_v^2v_{\rm th}^2(s)}. \nonumber
\end{eqnarray}

\begin{figure}[H]
	\centering
	\begin{subfigure}[c]{0.49\columnwidth}
	    \caption{Standard \label{fig:countcsi_std128_short}}
	    \centering
		\includegraphics[width=1.0\linewidth]{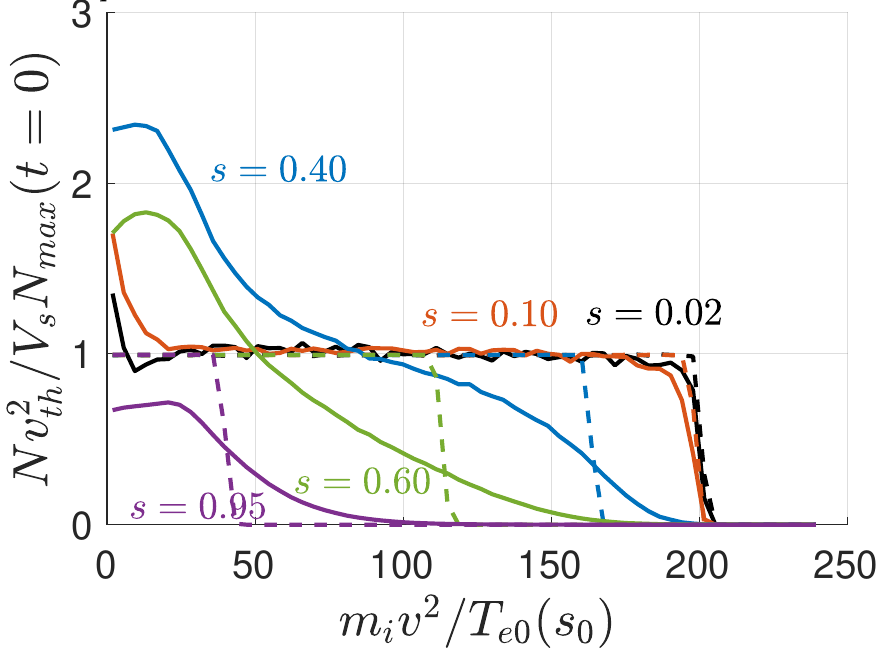}
	\end{subfigure}
	\begin{subfigure}[c]{0.49\columnwidth}
	    \caption{Adaptive \label{fig:countcsi_adp128_short}}
	    \centering
		\includegraphics[width=1.0\linewidth]{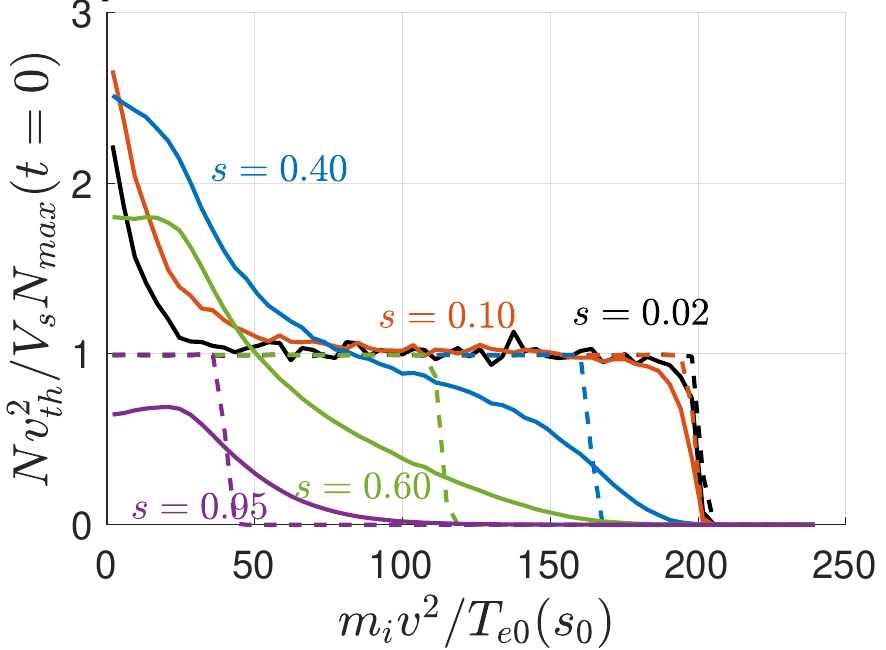}
	\end{subfigure}
	\caption{Ion marker count for different $s$-cuts for the (a) standard and (b) adaptive cases with $N_p=128$M at initial time (dashed) and at $c_st/a=612$ (solid). Here, $N_{max}(s,t=0)$ is the global maximum marker per $(s,v^2)$-bin count at initial time $t=0$.}
	\label{fig:countcsi}
\end{figure}

Figure~\ref{fig:countcsi} shows the various radial cuts of the ion marker number per $(s,v^2)$-bin corresponding to Figs.\ref{fig:dpvoli_std128_short} and \ref{fig:dpvoli_adp128_short} respectively, i.e. at the same instant $c_st/a=612$ and with $N_p=128$M. We note a general accumulation of markers towards lower energies. Near the magnetic axis at $s=0.02,0.10$ (black,red) however, there is greater marker accumulation to lower energies for the adaptive case compared to the standard case, thus concluding that the under-sampling near the magnetic axis is the result of phase-space volume corruption. The smaller extent of the under-sampling in Fig.\ref{fig:dpvoli_adp128_short} for the adaptive case seems to be ensured by the larger accumulation of markers shown in Fig.~\ref{fig:countcsi_adp128_short}. Further investigation in a future work will be required to fully understand this issue.

The problem of under-sampling for the standard cases Figs.\ref{fig:dpvoli_std256_short}-\ref{fig:dpvoli_std128_short} and \ref{fig:countcsi_std128_short} thus explains the density drop in Fig.\ref{fig:densi_short} and temperature rise in Fig.\ref{fig:tempi_short}. The former is due to the lack of markers being accounted for, and the latter is due to the lack of the contribution of low-energy markers. As temperature is the average energy over density, the under-estimated density also contributes to the over-estimation of temperature. In terms of the gyrokinetic equations of motion (\ref{eq:eom}), the difference between the standard and adaptive schemes lies in the evaluation of the $E\times B$ drift term involving the self-consistent gyroaveraged electrostatic potential $\phi$. The adaptive scheme allows for a more accurate evaluation of the r.h.s.~of the QNE (\ref{eq:qne_hyb}) using a better control variate $f_0$, this in turn leading to a more accurate solution for $\phi$. Moreover, under the standard scheme, a better evaluation of $\phi$ requires a higher $N_p$. Indeed, based on Figs.\ref{fig:dpvoli_std256_short}-\ref{fig:dpvoli_std128_short}, the under-sampling issue near $s=0$ improves with higher $N_p$. Even though the incompressibility of phase space should be verified for any field $\phi$, be it self-consistent or not, non-smooth solutions of $\phi$ to (\ref{eq:qne_hyb}) as a result of poor (noisy) integration of the r.h.s. can lead to $E\times B$ drifts in the marker trajectories which are difficult to integrate in time thus violating phase space incompressibility. Though not shown, no such under-sampling is observed for the electrons in both the standard and adaptive cases.

\subsection{Late time profiles} \label{sec:late}
\begin{figure}[H]
	\centering
	\begin{subfigure}[c]{0.49\columnwidth}
	    \caption{Ions\label{fig:tempi}}
	    \centering
		\includegraphics[width=1.0\linewidth]{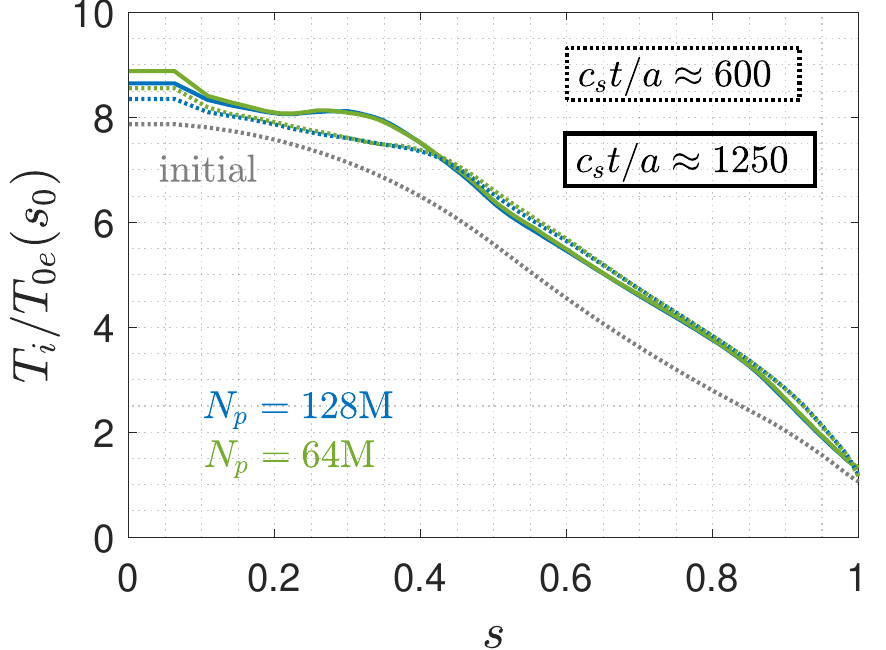}
	\end{subfigure}
	\begin{subfigure}[c]{0.49\columnwidth}
	    \caption{Electrons\label{fig:tempe}}
	    \centering
		\includegraphics[width=1.0\linewidth]{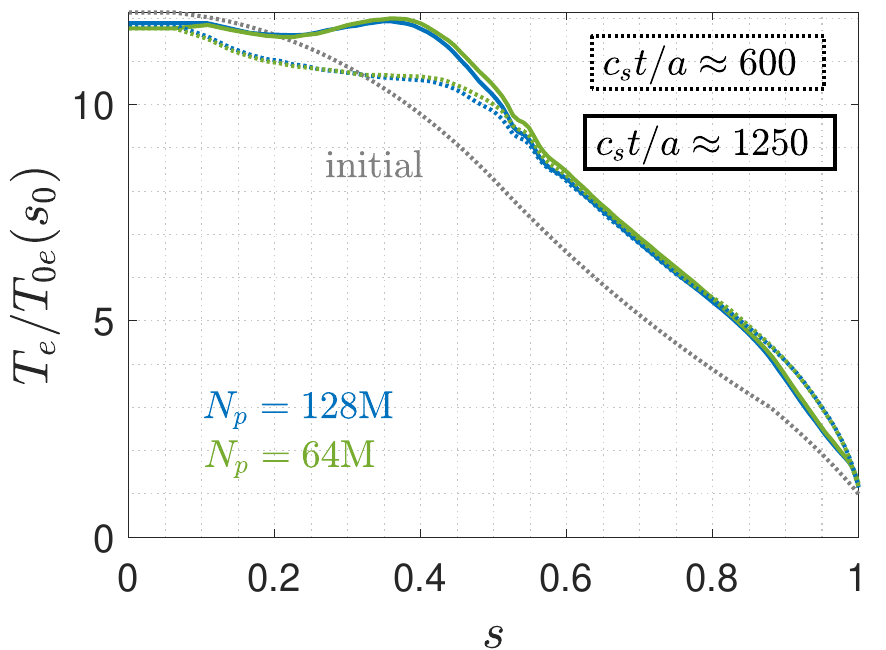}
	\end{subfigure}
	\caption{F.s.a.~gyrocenter temperature profiles time averaged over $c_st/a\in[1246,1266]$ (solid) and $c_st/a\in[592,613]$ (dotted) for various marker numbers $N_p$ under the adaptive scheme for (a) ions and (b) electrons. All colored curves show contributions by $f_0+\delta f$.}
	\label{fig:temp}
\end{figure}

\begin{figure}[H]
	\centering
	\begin{subfigure}[c]{0.9\columnwidth}
	    \caption{Ions \label{fig:RLTi}}
	    \centering
		\includegraphics[width=1.0\linewidth]{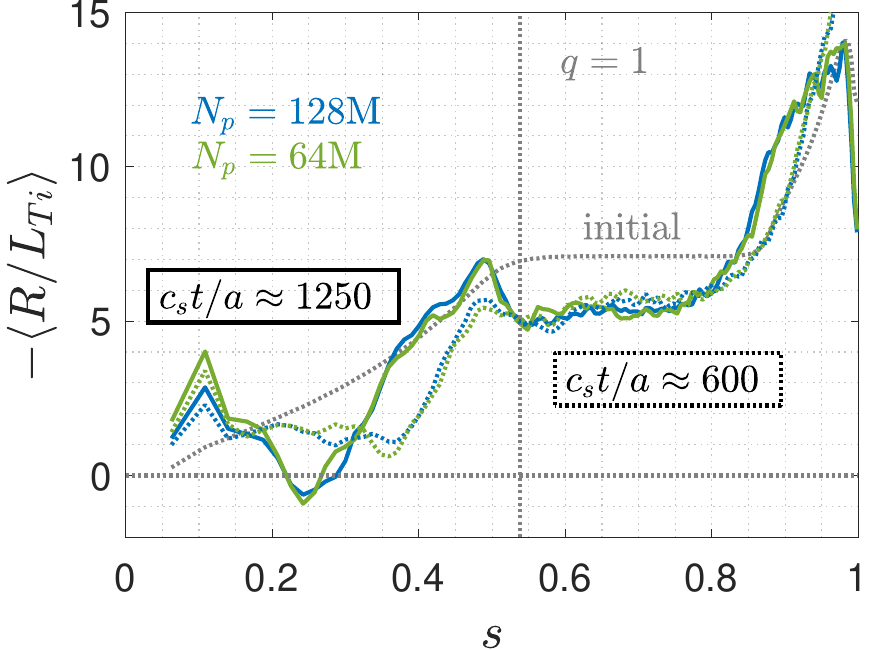}
	\end{subfigure}
	\begin{subfigure}[c]{0.9\columnwidth}
	    \caption{Electrons \label{fig:RLTe}}
	    \centering
		\includegraphics[width=1.0\linewidth]{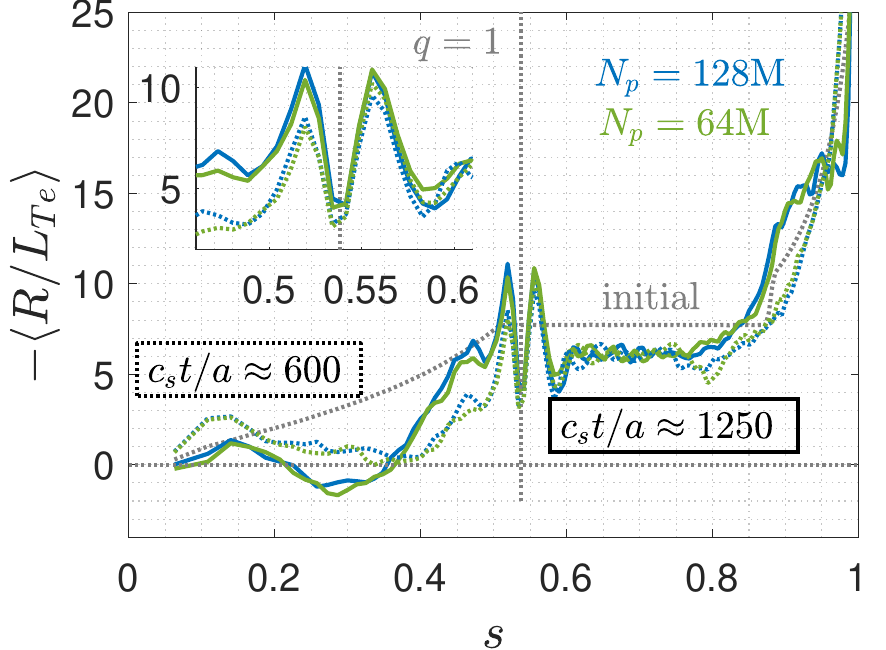}
	\end{subfigure}
	\caption{F.s.a.~gyrocenter temperature logarithmic gradient profiles time averaged over $c_st/a\in[1246,1266]$ (solid) and $c_st/a\in[592,613]$ (dotted), for various marker numbers $N_p$ under the adaptive scheme for (a) ions and (b) electrons. All colored curves show contributions by $f_0+\delta f$.}
	\label{fig:rlt}
\end{figure}

This section describes profiles for simulations that have been able to reach quasi-steady state with just one marker loading at initial time. For the considered heat sources, only adaptive cases clearly maintain adequate SNR levels to reach the simulation time $c_st/a=1266$ as shown in Fig.\ref{fig:snr}. Figure~\ref{fig:temp} shows the temperature profiles for the ions and electrons separately, contributed by the respective total distributions $f_0+\delta f$. Profiles contributed only by the background $f_0$ are not shown as they closely follow the total contribution, indicating that the adaptation rates are large enough and as a result $\fsa{\delta f}$ is very small. Also shown are the corresponding temperature profiles of the previous time window $c_st/a=[592,613]$ (colored dashed). We see that f.s.a.~temperature profile at $c_st/a=1266$ has further risen mainly around $s=0.4$, resulting in this region in a maximum deviation of about $20\%$ from initial values. As already mentioned for Fig.\ref{fig:tempi_short} the increase in temperature at the magnetic axis for the ions, Fig.\ref{fig:tempi}, is due to the under-sampling problem discussed in Sec.\ref{sec:pvol} and illustrated in Figs.\ref{fig:dpvoli_adp128_short}-\ref{fig:dpvoli_adp64_short}. Except for the temperature rise at the magnetic axis, the profiles with $N_p=128$M and $N_p=64$M of the respective species seem to be converged. The logarithmic gradients of these temperature profiles are shown in Fig.\ref{fig:rlt}. The risen peak in temperature around $s\in[0.3,0.4]$ has caused an increase and decrease of the $-\fsa{R/L_T}$ values to the left and right of this region, respectively. Furthermore, the temperature peak has led to negative values of $-\fsa{R/L_T}$. Both ions and electrons also have increased temperature gradients in the pedestal region $s\in[0.8,1.0]$. This is probably caused by edge buffer region and homogeneous Dirichlet condition at $s=1$ for $\phi$ both contribute to reduction of turbulence leading to profile steepening at the outer boundary. Logarithmic gradient values at the flat gradient region $s\in[0.6,0.9]$ on the other hand remains approximately constant in time after a decrease before the first time window $c_st/a\in[592,613]$. The corrugation of the electron temperature logarithmic gradient profile around $s\approx0.55$ has also remained relatively constant in amplitude.

\begin{figure}[H]
	\centering
	\begin{subfigure}[c]{0.9\columnwidth}
	    \caption{\label{fig:densi}}
	    \centering
		\includegraphics[width=1.0\linewidth]{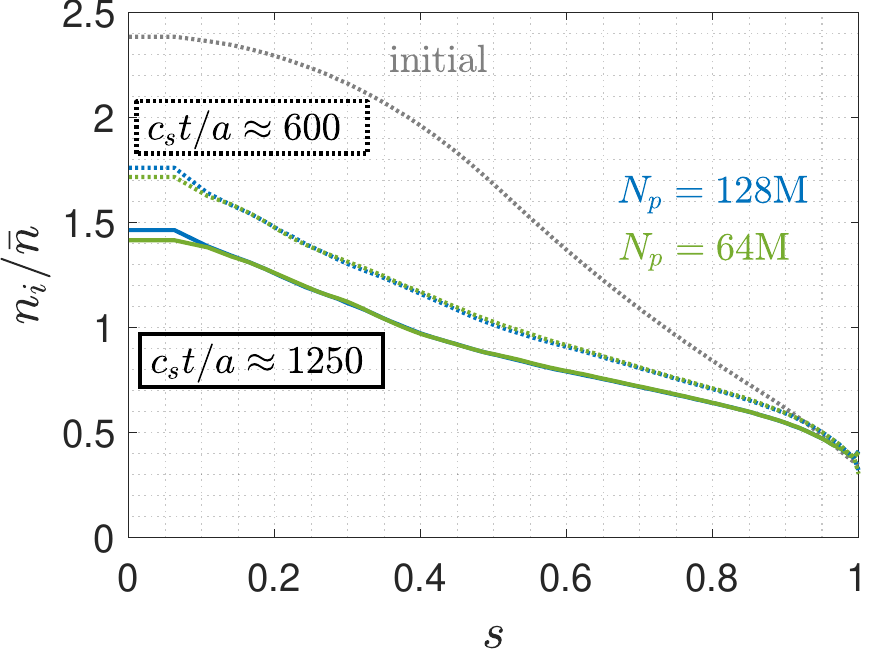}
	\end{subfigure}
	\begin{subfigure}[c]{0.9\columnwidth}
	    \caption{\label{fig:graddens}}
	    \centering
		\includegraphics[width=1.0\linewidth]{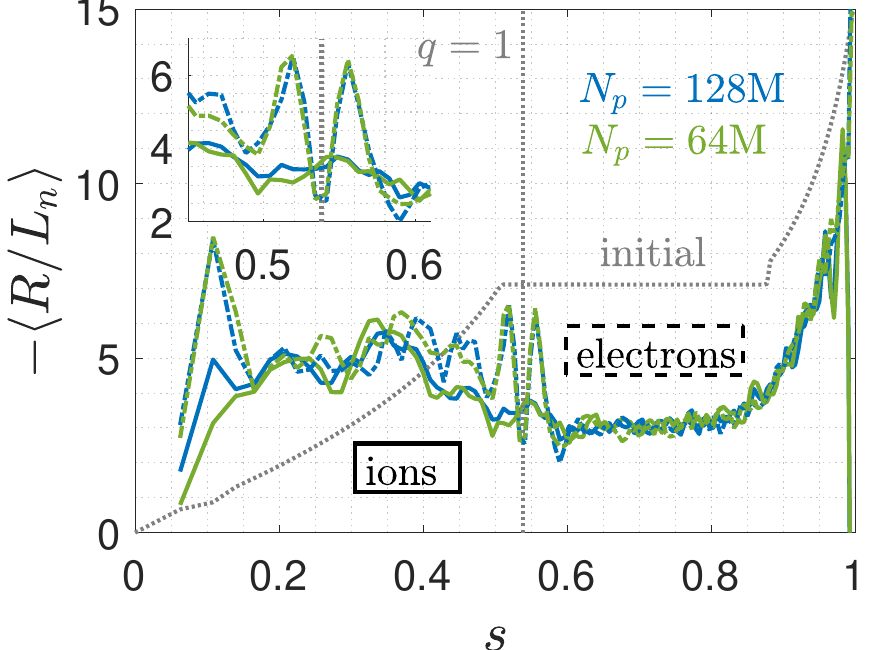}
	\end{subfigure}
	\caption{F.s.a.~(a) ion gyrocenter (solid) density and (b) ion gyrocenter (solid) and electron (dashed) density logarithmic gradient profiles time averaged over $c_st/a\in[1246,1266]$, for various marker numbers $N_p$ under the adaptive scheme. Also shown in (a) is the ion gyrocenter density in the previous time window $c_st/a\in[592,613]$ (dashed). All colored curves show contributions by $f_0+\delta f$.}
	\label{fig:dens}
\end{figure}

We now investigate the density profile evolution at quasi-steady state. Figure~\ref{fig:dens} shows the flux-surface- and time-average over $c_st/a\in[1246,1266]$ (colored solid) density and its logarithmic gradient profiles for the ions and electrons contributed by $f_0+\delta f$ under the adaptive scheme. For comparison, the density values at the previous time window $c_st/a\in[592,613]$ (colored dashed) are shown in Fig.\ref{fig:densi}. As the ion gyrocenter flux decreases approaching quasi-steady state, so does the change in ion gyrocenter density profiles given that the simulations include no particle sources, Fig.\ref{fig:densi}. The adaptive cases with $N_p=128$M and $64$M seem to have converged density profiles everywhere except near the magnetic axis, due to aforementioned ion under-sampling of phase-space. The difference in the profiles between the two simulations with different $N_p$ is however very small ($5\%$).Though not shown, the difference between ion gyrocenter and electron density for the adaptive cases (blue and green) remains approximately the same between the two time-averaging windows $c_st/a\in[592,613]$ and $c_st/a\in[1246,1266]$, see Fig.\ref{fig:ddens_short}. Taken as a whole, species densities have decreased by $50\%$ from their initial values. This amount of deviation certainly challenges the delta-f PIC constraint of $\|\delta f\|/\|f\|\ll 1$ in the standard approach without background adaptation. We consider now both ion gyrocenter (solid) and electron (dashed) density logarithmic gradient of the adaptive cases as shown in Fig.\ref{fig:graddens}. We first note the significant decrease of $50\%$ from their initial values in the radial region $s\in[0.4,1.0]$. For the region $s\in[0.2,0.4]$ however, there is a local increase in $-\fsa{R/L_n}$ as the actual density values decrease. We see once again the corrugation in the electron density logarithmic gradient near the $q=1$ flux surface which is absent for the ions, which has already been observed in Figs.\ref{fig:RLTe_short} and \ref{fig:RLTe} for electron temperature logarithmic gradient profiles.

\section{Conclusion} \label{sec:conclusion}

The aim of this work was to extend the capability of global gyrokinetic turbulence simulations to cases where strong deviations from the initial state occur. Such is typically the case in regions of strong gradients or for long flux-driven simulations. When a particle-based numerical approach is used, this requires to address the issue of accumulation of sampling noise, which was done in this work by introducing an adaptive f.s.a.~background as control variate. Specifically, the background that describes the gyrocenter distribution function assumes a Maxwellian form, with time-dependent density and temperature profiles. The main result of this work is the demonstration that the adaptive scheme allows for a reduction in computational cost by a factor as high as $8$ for obtaining a given numerical quality, in long, flux-driven simulations exhibiting strong deviations from the initial state.

To that end, an adaptive background density and temperature scheme was introduced. The background density and temperature was evolved through the ad-hoc relaxation equations (\ref{eq:relax_dens})-(\ref{eq:relax_ekin}) with associated relaxation rates. Along the simulation, the lower order f.s.a.~velocity moments of density and kinetic energy that tend to accumulate in the perturbed distribution function were kept low thanks to continuous transfer to the background Maxwellian. The adaptive scheme was implemented in toroidal geometry using the global gyrokinetic code ORB5~\cite{Lanti2020} to simulate mixed TEM-ITG regime turbulence. The electrons are modeled with an improved hybrid response~\cite{Idomura2016, Lanti2018, Lanti2019}, i.e.~when solving the Quasi-Neutrality-Equation (QNE) for the self-consistent electrostatic field the model takes into account the drift-kinetic of all electrons (trapped and untrapped) to the zonal perturbations, while for non-zonal perturbations trapped electrons still respond drift-kinetically, the passing electrons however, adiabatically. Heat source radial profiles for the ions and electrons, respectively, estimated based on previous temperature-gradient-driven runs were used as local fixed power sources for flux-driven runs presented in this work. The adaptive scheme used a canonical Maxwellian control variate Eq.(\ref{eq:f0cm}), and adapted both density and temperature profiles of ions and electrons independently. When imposing Eq.(\ref{eq:qnrhs_fsa}) the evolution of ion and electron f.s.a.~gyrocenter densities are however not independent. A comparison of two methods of calculating the r.h.s.~correction to the QNE, Eq.(\ref{eq:qne_corr}), was conducted and it was shown that the correction term is necessary to keep correct zonal flow structures.

When compared to the non-adaptive cases, results of the adaptive cases showed higher heat fluxes and lower zonal flow shearing rate amplitudes. The adaptive cases kept the SNR at quasi-steady values, with greatly reduced standard deviation of marker weights. This is demonstrated to be feasible even with adaptive cases having only a quarter of the number of markers used by the non-adaptive (standard) cases (from $256$M to $64$M markers). Only the simulations using the adaptive cases managed to reach quasi-steady state from a single marker loading at initial time. Phase-space volume diagnostics were used to detect phase-space volume depletion at low energies near the magnetic axis for ions. Though this problem occurred for both non-adaptive and adaptive cases, the latter is significantly less affected. Nonetheless, both schemes still suffer from growing regions in phase space that are potentially under-sampled as the simulations run over transport time scales due to marker diffusion at the velocity sampling boundary. For scenarios with even greater profile deviation, a resampling of markers is expected to become necessary.

As future work, investigations into the benefits of also adapting the background parallel flow could be conducted. Introducing time-dependence into background ion gyrocenter density allows one to partly de-linearize the ion polarization density term in the QNE. The implications of using this time-dependent density in the field equation could be explored. So as to also extend the evolving background approach to electromagnetic simulations, a similar correction term would need to be added to Amp\`{e}re's law as the one implemented in the QNE. This would then allow efficient flux-driven simulations of kinetic ballooning, tearing, and internal kink modes. In the presence of fast ions, a time-dependent background could also be useful when simulating in particular Alfv\'{e}n or energetic particle modes.

Further sophistication to the control variate could be envisaged. A control variate expanded in a set of basis functions could be pursued. Though only used as an offline diagnostic, Ref.\cite{Bottino2022} has expressed in ORB5 the background distribution as a polynomial expansion in the space of invariants of the unperturbed trajectories. Representing the background velocity distribution in terms of Hermite and Laguerre polynomials basis functions could be used in tandem with collision operators expressed in such a basis~\cite{Frei2022}. A mixed representation akin to the XGC code~\cite{Ku2009,Ku2016}, where the control variate consists of an analytic function plus a correction term represented on a phase-space grid could be an alternative. Nonetheless, complexity added to background translates to complexity added to the code as a whole. A highly complex control variate risks inheriting the disadvantages of both the particle- and grid-based approaches. It may be best to fall-back to the full-f PIC approach for simulations exhibiting high relative fluctuation amplitudes, i.e.~in particular for simulating edge or Scrape-Off-Layer (SOL) conditions.

\begin{acknowledgements}

The authors would like to thank Stephane Ethier and Alexey Mishchenko for their valuable discussions and inputs. This work is part of the EUROfusion `Theory, Simulation, Validation and Verification' (TSVV) Task. This work has been carried out within the framework of the EUROfusion Consortium, via the Euratom Research and Training Programme (Grant Agreement No 101052200 — EUROfusion) and funded by the Swiss State Secretariat for Education, Research and Innovation (SERI). Views and opinions expressed are however those of the authors only and do not necessarily reflect those of the European Union, the European Commission, or SERI. Neither the European Union nor the European Commission nor SERI can be held responsible for them. This work is also supported by a grant from the Swiss National Supercomputing Centre (CSCS) under project IDs s1232, s1252, and was partly supported by the Swiss National Science Foundation. 
\end{acknowledgements}

\section*{Data Availability Statement}

The data that support the findings of this study are available from the corresponding author upon reasonable request.



\section*{References}

\bibliography{paper2024_aip}

\end{document}